\documentstyle[citesort,epsf,a4,12pt]{article}

\def\lsim{\;\raise0.3ex\hbox{$<$\kern-0.75em\raise-1.1ex\hbox{$\sim$}}\;}
\def\gsim{\;\raise0.3ex\hbox{$>$\kern-0.75em\raise-1.1ex\hbox{$\sim$}}\;}
\def\ZZ{\hbox{\it Z\hskip -4.pt Z}} \newcommand{\ba}{\begin{array}}
\newcommand{\ea}{\end{array}} \newcommand{\bea}{\begin{eqnarray}}
\newcommand{\eea}{\end{eqnarray}} \newcommand{\nn}{\nonumber}

\begin{document}

\begin{titlepage}

\begin{flushright} Orsay LPTHE-97-68 \end{flushright}

\vspace{3cm} \centerline{\Large{\bf{Neutralino Cascades in the (M+1)SSM}}}

\vspace{2cm} \centerline{\bf{U. Ellwanger and C. Hugonie}}
\centerline{Laboratoire de Physique Th\'eorique et Hautes Energies}
\centerline{Universit\'e de Paris-Sud, F-91400 Orsay} \vspace{2cm}

\begin{abstract}

In the (M+1)SSM an additional gauge singlet Weyl spinor appears in the
neutralino sector. For a large part of the parameter space this approximative
eigenstate is the true LSP. Then most sparticle decays proceed via an
additional cascade involving the NLSP $\rightarrow$ LSP transition, where the
NLSP is the non-singlet next-to-lightest neutralino. We present a comprehensive
list of all processes, which contribute to the NLSP $\rightarrow$ LSP
transition, the partial widths and the total NLSP decay rate. We perform a scan
of the parameters of the model compatible with universal soft terms, and find
that the NLSP life time can be quite large, leading to macroscopically
displaced vertices. Our results imply that the signatures for sparticle
production in the (M+1)SSM can be very different from the MSSM, and are
important for calculations of the abundance of dark matter in this model.

\end{abstract}

\end{titlepage}

\section{Introduction}

The supersymmetric extension of the standard model with an additional singlet
superfield \cite{UMC,NMSSM1,NMSSM2,Higgs,walls,neu2,RGE2,Steph} has some
attractive features: the superpotential can be chosen to be scale invariant,
hence the only dimensionful parameters -- and thus the electroweak scale --
enter via the soft supersymmetry breaking terms. With a scale-invariant
superpotential, and assuming universal soft terms at the GUT scale, the model
has the same number of free parameters as the MSSM. Several analyses of the
parameter space of the model have previously been performed in
\cite{NMSSM2,UMC}. It has been found that a considerable region is consistent
both with theoretical constraints (correct $SU(2)_L \times U(1)_Y$ symmetry
breaking, no squark or slepton vev's, neutral LSP) and experimental lower
bounds on sparticle and Higgs masses.

It is very important to investigate in what respect the phenomenology of the
(M+1)SSM differs from that of the MSSM. The signatures for sparticle production
could be different, and one would like to know which processes could serve to
distinguish the two models.

The particle content of the (M+1)SSM differs from the MSSM in the form of
additional gauge singlet states in the Higgs sector (1 neutral CP-even and 1
CP-odd state) and in the neutralino sector (a two component Weyl fermion).
These states are mixed with the corresponding ones of the MSSM, and the
physical states have to be obtained from the diagonalization of the mass
matrices in the corresponding sectors. An interesting result of the analyses in
\cite{NMSSM2,UMC} is that, for most of the parameter space, the mixing angles
involving the singlet states are actually quite small. Consequently there exist
physical {\it quasi singlet} states which have only small couplings to the
gauge bosons and the MSSM sparticles such as charginos, squarks and sleptons.
These states then have only small production cross sections and it seems to be
nearly impossible to observe them in present or future experiments.

A notable exception can occur, however, in the neutralino sector. In the MSSM
the neutralino sector consists of two gauginos (the bino and the neutral wino)
and two higgsinos. Typically the LSP -- the lightest supersymmetric particle,
which is stable if one assumes, as we do, R-parity conservation -- is the
lightest eigenstate of the neutralino mass matrix. The LSP will appear as one
of the final states of each sparticle decay, and its non-observability is
responsible for the well-known missing energy/momentum signature of sparticle
production.

The situation in the (M+1)SSM, where a singlino state is added to the
neutralino sector, depends crucially on its mass with respect to the MSSM LSP
mass: If the singlino is heavier, it will very rarely be produced and it will
be practically unobservable. If it is lighter (and is thus the true LSP), it
will now appear at the end of the decay chain of sparticles decays. To be more
specific, from the analyses performed in \cite{UMC} one finds that the MSSM
LSP, within the allowed parameter space of the (M+1)SSM, is essentially a bino.
In the singlino LSP case of the (M+1)SSM one has to keep in mind the small
couplings of the singlino to all the other particles. If the sparticles are
heavier than the bino (which turns out to be always the case, except for some
large supersymmetry breaking terms that yield sparticles out of reach for LEP2)
they thus prefer to decay into the bino to which they couple more strongly.
Only then the bino will decay into the singlino LSP, which will give rise to an
additional cascade in the sparticle decay chain. Since this process modifies
the signatures for sparticle production considerably, we will investigate the
bino to singlino transition in detail in this paper. 

Production and decay of neutralinos have previously been discussed in the MSSM
in, e.g., ref. \cite{neu1} and in the (M+1)SSM in ref. \cite{neu2}. (Many of
the formulas of the partial decay widths in our appendix D can be found in
these papers.) In ref. \cite{neu2} production cross sections and branching
ratios  of neutralinos in the (M+1)SSM have been presented for several
scenarios  concerning the low energy parameters. 

Here, however, we are interested in a comprehensive analysis of the part of the
parameter space of the (M+1)SSM which is compatible with universal soft terms
at the GUT scale, which corresponds to the case of a singlino LSP, and where
sparticle production is kinematically possible at LEP2. Our aim is to see,
which bino lifetimes and bino branching ratios are possible under these
assumptions. With our results at hand one can decide, which signatures for
sparticle production (beyond the ones of the MSSM) are most promising in the
framework of the (M+1)SSM, and which part of the parameter space can be
tested. 

Our approach follows closely the one of \cite{UMC}, up to slightly different
experimental constraints on the parameter space: We start to scan the complete
parameter space of the model (the universal soft terms and Yukawa couplings,
see the next section) as defined at the GUT scale. For each point in the
parameter space we compute the effective parameters at low energy by
integrating the renormalization group equations from $M_{GUT}$ down to $M_Z$.
Then, for each set of parameters, we minimize numerically the effective
potential including the one loop radiative corrections induced by top quark and
squark loops. We check whether the absolute minimum of the potential breaks
$SU(2)_L \times U(1)_Y$ as desired, whether squarks or sleptons do not assume
vev's, and whether the physical masses of the top quark and the sparticles
satisfy the (model independent) present experimental constraints. (Details are
given in the next section.) Finally we require that the LSP,
$\tilde{\chi}^0_1$, is essentially a singlino state (otherwise the signatures
for sparticle production are not different from the MSSM), and that the mass of
the NLSP $\tilde{\chi}_2^0$ (which is the bino) is below $M_Z$. Under this
approximate condition sparticle production at LEP2 is kinematically allowed.

For each of the $\sim 10^4$ remaining points in the parameter space, we compute
the following decay widths: $\tilde{\chi}_2^0 \to \tilde{\chi}_1^0 l^+ l^-$,
$\tilde{\chi}_2^0 \to \tilde{\chi}_1^0 \nu \bar{\nu}$, $\tilde{\chi}_2^0 \to
\tilde{\chi}_1^0 q \bar{q}$ (in all cases we take $Z$, Higgs and slepton or
squark exchange into account), $\tilde{\chi}_2^0 \to \tilde{\chi}_1^0 +$Higgs
and $\tilde{\chi}_2^0 \to \tilde{\chi}_1^0 + \gamma$. (The radiative decays
into a photon have previously been considered in the MSSM in, e.g., refs.
\cite{HaWy,rad}, and in the (M+1)SSM in ref. \cite{neu2}.) The results give us
both the life time and the branching ratios of the \linebreak  NLSP $\to$ LSP
transition for each point in the parameter space of the model, which is
consistent with universal soft terms, present experimental constraints, and
which is of potential phenomenological relevance for LEP2. 

Clearly many steps of this procedure (e.g. the integration of the RGEs, the
minimization of the effective potential, the diagonalization of the mass
matrices and the phase space integrals) require numerical methods. These allow
us, however, to obtain the results with satisfactory accuracy. On the other
hand, we find it very desirable to understand at least the rough features of
our results (and of the range of the low energy parameters) using analytic
approximations to the integrated RGEs, the minimization of the effective
potential, and the diagonalization of the mass matrices. Therefore we spend
some time in sect. 3 in order to discuss the interplay between the different
theoretical and experimental constraints on the parameters within such analytic
approximations. These approximations allow us to understand the relative
importance and the orders of magnitude of the different decay widths in sect.
4. The results on the different decay widths, branching ratios and the total
life time presented in sect. 4 (and in the figures) are, however, based on the
"exact" numerical procedure.

Our results show that, even in the limit of tiny couplings of the singlino, a
priori a large number of different processes can contribute to the bino to
singlino transition. Only after a detailed investigation of all the partial
widths we find that only a few of them are relevant: Essentially the three body
decays with two leptons in the final state (via virtual slepton exchange) or
with $q \bar{q}$ in the final state (via virtual $Z$ exchange), and in some
cases the two body decay into a real singlet Higgs scalar or a photon.
Interestingly enough we find that, for small enough Yukawa couplings, the
lifetime of the bino becomes so large that displaced vertices appear to be
visible.

Two cosmological issues should also be discussed within the (M+1)SSM, namely
domain walls and dark matter. The (M+1)SSM with a scale invariant
superpotential has a discrete $\ZZ_3$ symmetry, which can lead to the formation
of domain walls with an unacceptable energy density during the electroweak
phase transition \cite{walls}. As discussed in \cite{walls}, possible ways out
of this problem are to embed the discrete symmetry into a gauge symmetry at
some large scale, or to add tiny mass terms, which do not modify the
phenomenology in a visible way, but which break the $\ZZ_3$ symmetry
sufficiently such that the domain walls are removed.

The LSP of any supersymmetric theory with conserved R parity is a priori a
welcome candidate for cold dark matter. It will necessarily be produced in
sparticle decays in the early universe, and its relic density will strongly
depend on its annihilation cross section. The (M+1)SSM has been considered in
this respect in \cite{Steph}, where both upper and lower limits on the LSP
relic density have been imposed. In \cite{Steph} it has been argued that the
singlino LSP scenario of the (M+1)SSM is essentially ruled out, since the pair
annihilation cross section is too small, and consequently the relic density is
too large. However, in \cite{Steph} only the LSP pair annihilation has been
considered. In particular in the case of small Yukawa couplings the situation
for a singlino LSP is, however, much more complicated: The binos could pair
annihilate before the LSP is produced, and the bino-singlino coannihilation
rate is much larger than the singlino pair annihilation rate. In order to
determine the dark matter constraints in the (M+1)SSM with singlino LSP
reliably it is thus absolutely necessary to know the bino lifetime or the bino
to singlino decay rate. Apart from the modified signatures for sparticle
production the results of this paper will thus also find applications in the
investigation of the dark matter in the (M+1)SSM.

The paper is organized as follows: In the next section we present the
lagrangian and discuss briefly the method of the scanning of the acceptable
parameter space; this procedure follows the one of ref.~\cite{UMC}. In section
three we study the range of parameters in the singlino LSP scenario in some
detail, focussing on analytic approximations. In section four we investigate
all possible bino to singlino decay channels, the corresponding contributions
to the partial bino decay widths, and the bino lifetime. We present both
approximate analytic results, and "exact" results based on the numerical
procedure. In section five we discuss our results and its phenomenological
consequences.

\section{Parameter Space of the (M+1)SSM} \label{secparam}

In this section, we study the parameter space of the model with the same
assumptions as in \cite{UMC}. The superpotential of the (M+1)SSM is given by
\bea W = \lambda H_1 H_2 S + \frac{1}{3} \kappa S^3 + \ldots \label{sp} \eea
where the ellipsis stand for quarks and leptons Yukawa couplings, \bea H_1 =
\left( \ba{c} H_1^+ \\ H_1^0 \ea \right), \quad H_2 = \left( \ba{c} H_2^0 \\
H_2^- \ea \right) \quad \mbox{and} \quad H_1H_2 = H_1^+H_2^- - H_1^0H_2^0 .
\eea

Here the Higgs doublet $H_1$ couples to the up-type quarks, and $H_2$ to the
down-type quarks and the charged leptons. Therefore the usual parameter $\beta$
is given by \bea \tan\beta = \frac{h_1}{h_2} \eea with $h_i = \langle H_i^0
\rangle$. $S$ denotes the gauge singlet superfield beyond the MSSM. The
superpotential contains no $\mu H_1 H_2$ term. An effective $\mu$ term is
generated once the scalar component of the singlet $S$ acquires a vev $s$: 
\bea \mu = \lambda s \label{mu} . \eea

The only dimensionful parameters of the model are the supersymmetry breaking
gaugino masses, scalar masses and trilinear couplings (for simplicity we do not
display the terms involving squarks or sleptons): \bea {\cal L}_{soft} & = &
\frac{1}{2} \left( M_3 \lambda_3^a \lambda_3^a + M_2 \lambda_2^i \lambda_2^i +
M_1 \lambda_1 \lambda_1 \right) + \mbox{ h.c.} \nn \\ & & - m_1^2 |H_1|^2 -
m_2^2 |H_2|^2 - m_S^2 |S|^2 \nn \\ & & - \lambda A_{\lambda} H_1 H_2 S -
\frac{1}{3} \kappa A_{\kappa} S^3 + \mbox{ h.c.} \label{Lsoft} \eea
$\lambda_1$, $\lambda_2$ and $\lambda_3$ are the gauginos of the $U(1)_Y$,
$SU(2)_L$ and $SU(3)_c$ gauge groups respectively. The scalar components of the
Higgs in (\ref{Lsoft}) are denoted by the same letters as the corresponding
chiral superfields.

The scalar potential contains the standard F and D terms, the supersymmetry
breaking terms and one loop radiative corrections of the form \bea V_{rad} =
\frac{1}{64\pi^2} \mbox{STr} {\cal M}^4 \ln \left( \frac{{\cal M}^2}{Q^2}
\right) \label{Vrad} . \eea

In (\ref{Vrad}) we take into account only top quark and squark loops, but we
include the numerically important contributions beyond the leading log
approximation which result from the complete top squark mass matrix. $Q^2$
denotes the renormalization point, and all the parameters in eqs.~(\ref{sp}),
(\ref{Lsoft}) and (\ref{Vrad}) have to be taken at the scale $Q^2 \sim M_Z^2$.

The supersymmetry breaking terms of the model are constrained by requiring
universal terms at the scale $M_{GUT} \sim 10^{16}$~GeV. The independent
parameters of the model are thus universal gaugino masses $M_0$ (always
positive in our convention), a universal mass for the scalars $m_0^2$, a
universal trilinear coupling $A_0$ (either positive or negative), and the
Yukawa couplings $\lambda_0$ and $\kappa_0$ of the superpotential (\ref{sp}) at
the scale $M_{GUT}$. In addition the top quark Yukawa coupling affects the
renormalization group evolution of the parameters from $M_{GUT}$ down to the
electroweak scale. The value of the $Z$ mass fixes one of these parameters with
respect to the others, so that we end up with 5 free parameters at the GUT
scale, as many as in the MSSM with universal soft terms.

Following the same procedure as in \cite{UMC}, we perform a scan over the
complete parameter space of the model at $M_{GUT}$, integrate the
renormalization group equations (RGE) down to the electroweak scale, and
minimize the low energy effective potential including the radiative corrections
(\ref{Vrad}) numerically in each case. We check whether we have found the
absolute minimum of the potential, and verify whether squarks or sleptons do
not assume vev's, which would break color and/or electromagnetism. Already at
this stage, the condition to avoid selectron vev's (which are the most
dangerous ones) yields a constraint on the parameter space \cite{UMC}: \bea
\frac{A_0}{M_0} \gsim -2.5 \label{A0M0} \eea In the remaining cases we
diagonalize numerically the mass matrices, compute the physical masses of all
particles and impose the following experimental constraints: \bea
m_{\tilde{\nu}} & > & M_Z/2 \: \mbox{GeV} \quad \cite{exp1} , \nn \\ m_t & = &
175 \pm 6 \: \mbox{GeV} \quad \cite{exp2} . \label{excon} \eea

Note that, since signatures for sparticles production in the present scenario
may be different from the MSSM, we cannot apply the standard MSSM analysis to
the latest data from LEP1.5 and LEP2. This data should rather be reanalysed, in
the context of the (M+1)SSM, using the results of the present paper. However,
the LEP1 results on the $Z$ width and thus the sneutrino mass $m_{\tilde{\nu}}$
remain valid. Moreover it turns out that the essential properties of the
neutralino sector do not depend on the details of the lower limits on, e.g.,
the chargino or slepton masses. 

Furthermore, within the present assumption of universality of the soft
parameters at the GUT scale and the singlino LSP scenario, eqs.~(\ref{excon})
imply already strong constraints on the other new particle masses (cf
section~\ref{secSing}), so that nearly all the other experimental bounds turn
out to be automatically satisfied: \bea m_{\chi^0_i} + m_{\chi^0_j} > M_Z & or
& \Gamma(Z \rightarrow \tilde{\chi}^0_i \tilde{\chi}^0_j) < \ 7 \: \mbox{MeV}
\quad \mbox{if} \quad (i,j) = (1,1) , \nn \\ & & \Gamma(Z \rightarrow
\tilde{\chi}^0_i \tilde{\chi}^0_j) < 30 \: \mbox{MeV} \quad \mbox{if} \quad
(i,j) \neq (1,1) , \nn \\ m_{H^\pm} > 150 \: \mbox{GeV} & , & m_{A^0} > 130 \:
\mbox{GeV} , \nn \\ m_{\tilde{t}_1} > 190 \: \mbox{GeV} & , & m_{\tilde{g}} >
280 \: \mbox{GeV} , \nn \\ m_{\chi_1^\pm} > 60 \: \mbox{GeV} & , &
m_{\tilde{l}_R} > 60 \: \mbox{GeV} \nn \eea where $A^0$ is the lightest
non-singlet neutral CP-odd Higgs. The lightest non-singlet CP-even Higgs is in
the range from 100~GeV up to 140~GeV.

As emphasized in \cite{NMSSM2,UMC}, the allowed parameter space of the (M+1)SSM
is in general characterized by small values of the Yukawa couplings $\lambda$
and $\kappa$\linebreak ($\lambda,\kappa\lsim$~0.3). As we will see in the next
section the singlino LSP case corresponds to even smaller values of the Yukawa
couplings $\lambda , \kappa \lsim 10^{-2}$.

\section{Singlino LSP Scenario} \label{secSing}

In this section we present some special features of the singlino LSP scenario.
In particular, we derive some approximative constraints on the high energy free
parameters and some analytic approximations for the low energy masses and
mixing factors. These approximations are useful to understand the features of
our numerical results and will provide us with helpful guidelines for the
calculations of the next section. To this end we use approximate relations
between the low energy and high energy parameters of the model (as obtained
from the RGEs), and relations obtained from the minimization of the tree level
potential. At the beginning we assume that the Yukawa couplings $\lambda$ and
$\kappa$ are small, but this assumption will be justified below.

In order to derive the constraints on the parameters implied by the singlino
LSP scenario we first have to find approximate expressions for the singlino and
the lightest non-singlet neutralino masses. 

Let us start with the singlino mass $M$. From the neutralino mass matrix
(\ref{neumm}) of Appendix~\ref{appneu} one finds that the mixing of the
singlino to the higgsinos is proportional to $\lambda$ and thus relatively
small. Hence, the singlino remains an almost pure state, and its mass is  \bea
M = 2 \kappa s. \label{msing} \eea Using the minimization of the tree level
potential the vev $s$ and hence $M$ can be related to the bare parameters of
the model: For small Yukawa couplings (and hence $s \gg h_1,h_2$) the
minimization equation (\ref{min3}) for the singlet becomes \bea s \simeq
-\frac{A_\kappa}{4\kappa} \left( 1 + \sqrt{1-\frac{8m_S^2} {A_\kappa^2}}
\right) . \label{svev} \eea $A_\kappa$ and $m_S$ being only slightly
renormalized between $M_{GUT}$ and $M_Z$ (cf Appendix~\ref{appRGE}), one
obtains the singlino mass in terms of the GUT parameters: \bea M \simeq
-\frac{A_0}{2}\left( 1 + \sqrt{1-\frac{8m_0^2}{A_0^2}} \right) . \eea Note that
$M$ has the sign opposite to $A_0$. The condition for the minimum (\ref{svev})
to be deeper than the symmetric one ($h_1=h_2=s=0$) reads \bea A_0^2 \gsim 9
m_0^2 , \label{vacsta} \eea so that \bea \frac{2}{3}|A_0| \lsim |M| \lsim |A_0|
. \label{singmass} \eea

Next, we estimate the masses of the lightest non-singlet neutralinos. The
higgsino effective mass term $\mu = \lambda s$ turns out to be quite large (see
below). Since the mixing terms between the gauginos and the higgsinos are at
most of  $O(M_Z/2\mu)$, one finds that, to a good approximation, the lightest
non-singlet neutralinos are the (nearly pure) bino and the wino. Their masses
$M_1$ and $M_2$ are related to $M_0$ as given in appendix \ref{appRGE}: \bea
M_1 \simeq .5 M_2 \simeq .41 M_0 . \label{gaugmass} \eea

The condition for the singlino to be the LSP is given by $M < M_1$, which,
combined with (\ref{singmass}) and (\ref{gaugmass}), yields \bea |A_0| \lsim .6
M_0 . \label{alm} \eea (Note that this condition is compatible with the
necessary condition for the absence of color and/or electromagnetism breaking
vevs, eq.~(\ref{A0M0}).) Eq.~(\ref{alm}) together with eq.~(\ref{vacsta})
implies that the singlino LSP scenario discards large $|A_0|/M_0$ and
$m_0^2/M_0^2$ ratios, and it just corresponds to a very natural "gaugino
dominated scenario", gaugino masses being the largest soft terms. Then, the
masses of all non-singlet sparticles can be expressed in terms of $M_0$ and are
therefore strongly correlated.

For later use it is convenient to introduce a parameter $\eta$, defined by the
ratio of the masses of the lightest and next-to-lightest neutralinos:  \bea
\eta = \frac{m_{\chi_1^0}}{m_{\chi_2^0}} \simeq \frac{M}{M_1} . \label{eta}
\eea Unlike in the MSSM, it is not fixed by universality constraints at the GUT
scale, but it is rather a free parameter varying from $-1$ to $+1$.
Eqs.~(\ref{singmass}) and (\ref{gaugmass}) allow us to express $\eta$ easily in
terms of the bare parameters $A_0$ and $M_0$: \bea \eta \sim -2 \frac{A_0}{M_0}
. \label{eta1} \eea

Next, we wish to estimate the higgsino effective mass term $\mu = \lambda s$,
and show that it is quite large. Below, the knowledge of $\mu$ will allow us to
relate $\eta$ to the Yukawa coupling $\lambda$.  First, the minimization of the
tree level potential with respect to $h_1$ and $h_2$ gives the relation  \bea
\tan^2\beta = \frac{m_2^2+\mu^2+M_Z^2/2}{m_1^2+\mu^2+M_Z^2/2} . \label{tb} \eea
The condition for a non-trivial minimum corresponds to the condition that the
denominator of eq.~(\ref{tb}) has to be positive. The approximative solutions
of the RGEs (\ref{RGEm1}) and (\ref{RGEm2}) imply that $m_2^2 > 0$, whereas
$m_1^2 < 0$. Hence one obtains \bea \mu^2 \gsim - m_1^2 - .5 M_Z^2 \gsim 2.1
M_0^2 - .5 M_Z^2 . \label{muM01} \eea Since, in addition, phenomenological
constraints imply that $M_0$ is quite large ($M_0 \gsim$~110~GeV, see below),
one finds $\mu^2 \gg M_Z^2$. Actually, from our numerical results (within the
singlino LSP scenario) we find $\tan \beta \gsim 6$ which implies \bea \mu^2
\simeq 2.5 M_0^2 -.5 M_Z^2 . \label{muM0} \eea in agreement with (\ref{muM01}).

Next, we derive an upper limit on the Yukawa couplings within the singlino LSP
scenario. First, from the absence of a deeper unphysical minimum of the Higgs
potential with $h_2 = s = 0$, the following inequality can be derived
\cite{UMC}: \bea \kappa < 3.10^{-2} \frac{A_0^2}{M_0^2} . \label{kaplim} \eea
Furthermore, from the numerical analysis, the Yukawa couplings $\lambda$ and
$\kappa$ turn out to be closely related: \bea \kappa \sim \lambda^{1.5 \pm .3}
. \label{kappa}  \eea Using (\ref{eta1}) in (\ref{kaplim}) and (\ref{kappa})
one finds that the singlino LSP scenario requires small Yukawa couplings,
$\lambda , \kappa \lsim 10^{-2}$. 

On the other hand, no lower limit on the Yukawa couplings has been found in our
analysis; we allowed for couplings as small as $\lambda = 10^{-6}$. In this
regime one can show that the singlino LSP scenario follows automatically: Using
(\ref{msing}), (\ref{mu}),  (\ref{kappa}) and (\ref{muM0}), one gets \bea |M| =
2 \kappa |s| = 2 \kappa |\mu|/\lambda \sim 2 \lambda^{.5 \pm .3} | \mu | \sim
3.2 \lambda^{.5 \pm .3} M_0 , \label{Mlambda} \eea so that \bea | \eta | \sim
7.7 \lambda^{.5 \pm .3} . \label{eta2} \eea Hence, for very small values of
$\lambda$ ($\lambda \lsim 10^{-5}$ in our numerical analysis), the singlino LSP
scenario is always realized and $\eta \ll 1$. The compatibility of
eq.~(\ref{eta2}) with eq.~(\ref{eta1}) requires some relation between  the bare
parameters $A_0$, $M_0$ and $\lambda_0$ of the model which is, however, not
very stringent. 

Herewith we conclude the discussion of the constraints on the parameters of the
model implied by the singlino LSP scenario. With the help of these results we
can now obtain approximate expressions for all quantities which are required in
order to calculate the bino to singlino decay widths: the mixing parameters of
the singlino with the other non-singlet neutralinos, and the masses of the
sfermions and the Higgs bosons. Let us first study the mixing parameters:

For $| \eta |$ not too close to 1, one can expand the singlino and the bino
eigenstates in small mixing parameters in the basis of eq.~(\ref{basis}): \bea
N_{1i} & \sim & \left( \frac{\lambda g_2(h_2^2 - h_1^2)\mu}{\sqrt{2}(M -
M_2)(M^2 - \mu^2)} , \frac{\lambda g_1(h_1^2 - h_2^2)\mu}{\sqrt{2}(M - M_1)(M^2
- \mu^2)} , \right. \nn \\ & & \left. \frac{\lambda(\mu h_1 - Mh_2)}{M^2 -
\mu^2} , \frac{\lambda(\mu h_2 - Mh_1)}{M^2 - \mu^2} , 1 \right) , \label{N1i}
\\ N_{2i} & \sim & \left( \frac{-g_1g_2(M_1(h_1^2 + h_2^2) + 2\mu
h_1h_2)}{2(M_2 - M_1)(M_1^2 - \mu^2)} , 1 , \frac{g_1(\mu h_2 +
M_1h_1)}{\sqrt{2}(M_1^2 - \mu^2)} , \right. \nn \\ & & \left. \frac{-g_1(M_1
h_2 + \mu h_1)}{\sqrt{2}(M_1^2 - \mu^2)} , \frac{\lambda g_1(h_1^2 -
h_2^2)\mu}{\sqrt{2}(M_1 - M)(M_1^2 - \mu^2)} \right) . \label{N2i} \eea Using
eqs.~(\ref{gaugmass}), (\ref{muM0}), (\ref{eta}) and $h_1 \gg h_2$, these
components can be expressed in terms of $\eta$, $M_1$ and $M_Z$. (However,
(\ref{N1i}) and (\ref{N2i}) are not valid anymore in the degenerate case
$|\eta| \rightarrow 1$.) These expressions for the mixing parameters will be
used extensively for the analytic approximations in the next section.

Next, we turn to the sfermion sector. The lightest states are the sneutrinos
$\tilde{\nu}$ and the "right handed" charged sleptons $\tilde{l}_R$. The
approximate expressions for their masses are (cf Appendix~\ref{appRGE}) \bea
m_{\tilde{l}_R}^2 = & m_E^2 - \sin^2\theta_W M_Z^2 \cos 2\beta & \simeq .15
M_0^2 + .23 M_Z^2 , \label{msel} \\ m_{\tilde{\nu}}^2 = & m_L^2 +
\frac{1}{2}M_Z^2 \cos 2\beta & \simeq .52M_0^2 - .5 M_Z^2 . \label{mnu} \eea
The lower limit on $m_{\tilde{\nu}}$ (\ref{excon}) combined with
eq.~(\ref{mnu}) gives a lower limit on $M_0$ ($\gsim$~110~GeV), which in turn
puts a lower limit on $m_{\tilde{l}_R}$ ($\gsim$~60~GeV). "Left handed" charged
sleptons and squarks are always much heavier, hence uninteresting for the
present phenomenology.

The lower limit on $M_0$ also yields a lower limit on the bino mass:\linebreak
$m_{\chi_2^0} \gsim$~30~GeV. Subsequently we restrict our analysis to the
regime $m_{\chi_2^0} < M_Z$, where sparticle production at LEP2 (at least the
pair production of binos) is kinematically allowed. In terms of $M_0$ this
corresponds to $M_0 \lsim$~230~GeV. In this region of the parameter space, the
bino is the NLSP. Note that for larger values of $M_0$ the D-term in
eq.~(\ref{msel}) becomes negligible and one has \bea m_{\tilde{l}_R}^2 \simeq
.15 M_0^2 < M_1^2 \simeq .17 M_0^2 . \label{slepbino} \eea Thus, for $M_0
\gsim$~320~GeV, $\tilde{l}_R$ turns out to be the NLSP. We shall come back to
the case $m_{\chi_2^0} > M_Z$ in the last section.

The large value of $\mu$ implies that the lightest chargino $\chi_1^{\pm}$ is
mainly a wino of mass $M_2$, which is related to $M_0$ by eq.~(\ref{gaugmass}).
However, for small values of $M_0$ the higgsino component can be quite large
(up to 50\%) and $m_{\chi_1^{\pm}}$ smaller than $M_2$. The lower bound on
$M_0$ yields $m_{\chi_1^{\pm}} \gsim$~60~GeV.

Similarly, from the lower bound on $M_0$, we obtain the lower bound on the
gluino mass given in the previous section.

Finally, we briefly focus our attention on the Higgs sector. We see from the
mass matrices given in Appendix~\ref{apphgs} that the mixings of the CP-even
and CP-odd singlets with the non-singlet Higgs fields are proportional to
$\lambda$, hence small. Here again, the singlet sector decouples and we end up
with two almost pure singlet states, a scalar of mass \bea M_{S,33}^2 \simeq
\frac{1}{4} \sqrt{A_0^2-8m_0^2} \left( |A_0| + \sqrt{A_0^2-8m_0^2} \right) ,
\eea and a pseudoscalar of mass \bea M_{P,22}^2 \simeq \frac{3}{4} |A_0| \left(
|A_0| + \sqrt{A_0^2-8m_0^2} \right)  \eea where we have used eqs.~(\ref{ms33}),
(\ref{mp22}) and (\ref{svev}). The pseudoscalar is always heavier than the
scalar singlet and, using arguments similar to (\ref{Mlambda}), one finds that
their masses are both roughly proportional to $M \sim \lambda^{.5 \pm .3} M_0$
in the singlino LSP case. Therefore, the singlet states are the lightest Higgs
states.

In the non-singlet sector, we have one CP-odd pseudoscalar $A^0$ of mass \bea
M_{P,11}^2 \simeq \frac{0.3M_0^2}{\sin 2\beta} \eea where we have used
(\ref{muM0}) and the approximate solution of the RGE for $A_\lambda$
(\ref{RGEAl}) in the limit $m_0,A_0 \ll M_0$. As $\tan\beta$ is always quite
large, $A^0$ turns out to be relatively heavy as already remarked in
section~\ref{secparam}. The mixing term between the CP-even fields $H_{1R}$ and
$H_{2R}$ is proportional to: \bea \frac{M_{S,12}^2}{M_{S,22}^2-M_{S,11}^2} \sim
\cot \beta \ll 1 . \label{higgsmix} \eea $H_{1R}$ and $H_{2R}$ are then almost
pure states of mass \bea M_{S,11}^2 & \simeq & M_Z^2 ,\\ M_{S,22}^2 & \simeq &
0.3M_0^2\tan\beta > M_{S,11}^2 . \eea

These approximations must be taken with care, as they do not include the
numerically important radiative contributions beyond the RGE to the effective
potential (\ref{Vrad}). Nevertheless, we find from the numerical analysis (with
the radiative corrections to the effective potential included) the following
particle assignments and mass ranges in the Higgs sector, in agreement with our
rough estimates: \bea\ba{lclccccc} S_1 & \sim & S_R \qquad & 0 & \lsim &
m_{S_1} & \lsim & 60 \: \mbox{GeV}, \nn \\ S_2 & \sim & H_{1R} \qquad & 100 &
\lsim & m_{S_2} & \lsim & 120 \: \mbox{GeV}, \nn \\ S_3 & \sim & H_{2R} \qquad
& 130 & \lsim & m_{S_3} & \lsim & 360 \: \mbox{GeV}, \\ P_1 & \sim & S_I \qquad
& 0 & \lsim & m_{P_1} & \lsim & 130 \: \mbox{GeV}, \nn \\ P_2 & \sim & A^0
\qquad & 130 & \lsim & m_{P_2} & \lsim & 350 \: \mbox{GeV}. \nn \ea\eea
However, the upper bounds given above increase if one relaxes the condition
$M_1 < M_Z$, allowing for higher values of $M_0$.

\section{Bino to Singlino Transition}

In this section we compute the bino to singlino decay widths. As already
mentioned, this process is crucial as it will appear at the end of every
sparticle decay chain in the singlino LSP case. The different contributions are
shown in figs.~\ref{figexch}-\ref{figrad}. Exact formulae for the corresponding
decay widths are given in Appendix~\ref{appDW}. The production and decay of
neutralinos have already been studied in the (M+1)SSM framework for a few
selected points in the parameter space \cite{neu2}. Instead, we have computed
the partial and total decay widths numerically for each point in the parameter
space obtained from our numerical scanning. 

In the following, we first present some simple analytic approximations so as to
understand the main features of the bino to singlino transition. Then we
discuss our the "exact"  results, which are based on the numerical integration
of the RGEs, the numerical minimization of the full Higgs potential, the
numerical computation of the mixings and mass eigenvalues in the neutralino,
chargino and Higgs boson sector, and the integration of the exact phase space
integrals. These results turn out to be in good agreement with the analytic
approximations.

First of all, let us consider the tree level three body decay $\tilde{\chi}^0_2
\rightarrow \tilde{\chi}^0_1 f \bar{f}$ of fig.~\ref{figexch}. The fermions can
be charged leptons, neutrinos or quarks (in which case we end up with two
jets). All the decay widths are proportional to $\lambda^2$ -- one factor
$\lambda$ from the non-singlet component of the singlino, raised to the square
-- and are hence equally suppressed. Therefore, for each final state, we have
to check whether the virtual $Z$, sfermion or Higgs exchange gives the main
contribution and whether we have to compute interference terms.

Let us start with a pair of charged leptons in the final state. The partial
width via virtual $Z$ exchange is given by eq.~(\ref{DWZexch}) \cite{neu1}. It
depends on the mixing factor $O_{12}$ in the coupling $Z \tilde{\chi}^0_1
\tilde{\chi}^0_2$ defined by eq.~(\ref{O12}), and a phase space integral $I_Z$
defined by eq.~(\ref{IZ}). In our analytic approximations we assume a very
light singlino, i.e. $\eta$ small. (We shall come back later to the case $\eta
\not\rightarrow 0$.) Using (\ref{N1i}) and (\ref{N2i}) in the limit of large
$\tan \beta$ and $| \eta | \ll 1$, the mixing factor $O_{12}$ can be written
\bea O_{12} \simeq 1.7\,10^{-2} \lambda \left(\frac{M_Z}{M_1}\right)^2 . \eea
As we take $|\eta|$ small, the phase space integral $I_Z$ is of $O(10^{-1})$,
so that the decay width reads \bea \Gamma(\tilde{\chi}^0_2
\stackrel{Z}\longrightarrow \tilde{\chi}^0_1 l^+ l^-) \simeq 6.10^{-9}
\lambda^2 M_1 I_Z(\eta,\omega_Z)\sim 10^{-10} \lambda^2 M_1 . \eea

For the slepton exchange, the partial width is given by eq.~(\ref{DWsexch})
\cite{neu1}. Since the "left" type charged sleptons are always much heavier
than the "right" type ones, their contribution will be relatively unimportant.
The vertex factor involves the mixing factor $N_{12}$ defined in
eq.~(\ref{defn}). Using (\ref{N1i}) with the same assumptions as above yields
\bea \sqrt{2}\,g_1N_{12} \simeq .13 \lambda \left(\frac{M_Z}{M_1}\right)^2 .
\eea The partial width then reads \bea \Gamma(\tilde{\chi}^0_2
\stackrel{\tilde{l}_R}\longrightarrow \tilde{\chi}^0_1 l^+ l^-) \simeq
2.10^{-6} \lambda^2 M_1 \left(\frac{M_Z}{m_{\tilde{l}_R}}\right)^4
I_{\tilde{l}_R}(\eta,\omega_{\tilde{l}_R}) . \label{above1} \eea As for the $Z$
exchange, the phase space integral $I_{\tilde{l}_R}$, given by eq.~(\ref{Is}),
is of $O(10^{-1})$. One can infer from (\ref{msel}) that the ratio
$M_Z/m_{\tilde{l}_R}$ is always of $O(1)$. Eq.~(\ref{above1}) then gives \bea
\Gamma(\tilde{\chi}^0_2 \stackrel{\tilde{l}_R}\longrightarrow \tilde{\chi}^0_1
l^+ l^-) & \sim & 10^{-7} \lambda^2 M_1 \label{lepslep} \\ & \gg &
\Gamma(\tilde{\chi}^0_2 \stackrel{Z}\longrightarrow \tilde{\chi}^0_1 l^+ l^-)
\nn \eea

In the case of virtual Higgs exchange, the partial width is given by
eq.~(\ref{DWHexch}) and depends, in this case, on the mixing factors $Q_{a12}$
and $Q_{al}$ defined in eqs.~(\ref{Qa12}), (\ref{Qafd}), respectively. (Here
and below the index $l$ denotes a charged lepton and replaces the index $f$ in
eq.~(\ref{Qafd}).) First, we observe that if the lightest Higgs scalar (which
is the singlet) is too heavy to be produced on shell, the partial width for its
virtual exchange is proportional to $\lambda^6$, and hence completely
negligible: \bea Q_{112} \sim \lambda^2 \quad \mbox{and} \quad Q_{1l} \sim
\lambda \qquad \Longrightarrow \qquad \Gamma(\tilde{\chi}^0_2
\stackrel{S_R}\longrightarrow \tilde{\chi}^0_1 l^+ l^-) \sim \lambda^6 .
\label{above2} \eea The result is similar for the singlet pseudoscalar $S_I$,
which is always heavier than $S_R$. As shown in the previous section, the
second scalar $S_2$ is mainly $H_{1R}$, so that \bea & & Q_{212} \simeq
\frac{\lambda}{\sqrt{2}}N_{24} - \frac{g_1}{2}N_{13} \simeq \frac{\lambda g_1
h_1}{\mu} , \\ & & Q_{2l} \simeq \frac{m_lS_{22}}{\sqrt{2}h_2} \simeq
\frac{m_l}{\sqrt{2}h_1} \eea where $m_l$ denotes the lepton mass and we have
used eqs.~(\ref{N1i}), (\ref{N2i}) and (\ref{higgsmix}). Eq.~(\ref{DWHexch})
then gives \bea \Gamma(\tilde{\chi}^0_2 \stackrel{H_{1R}}\longrightarrow
\tilde{\chi}^0_1 l^+ l^-) \simeq 10^{-5} \lambda^2
\frac{M_1^3m_l^2}{m_{H_{1R}}^4} I_2(\eta,\omega_2) . \eea As before, the phase
space integral $I_2$, given by eq.~(\ref{Ia}), is of $O(10^{-1})$. The only
leptonic final state with sizable couplings to the Higgses is the
$\tau^+\tau^-$ pair. Taking for $H_{1R}$ a mass of order 100~GeV, we get \bea
\Gamma(\tilde{\chi}^0_2 \stackrel{H_{1R}}\longrightarrow \tilde{\chi}^0_1
\tau^+ \tau^-) \sim 10^{-14} \lambda^2 \frac{M_1^3}{(1 \mbox{GeV})^2} . \eea
Even if one takes $M_1 = M_Z$, this is completely negligible compared to
(\ref{lepslep}). The second pseudoscalar $A^0$ and the third scalar $H_{2R}$
being much heavier than $H_{1R}$ will give even smaller contributions.

To summarize, the dominant contribution to the $\tilde{\chi}^0_2 \rightarrow
\tilde{\chi}^0_1 l^+ l^-$ transition is the slepton exchange, and we do not
need to compute any interference term between diagrams. This remains valid for
any value of $\eta$, although the partial width can become significantly
smaller than (\ref{lepslep}) as the phase space is reduced for $|\eta|
\rightarrow 1$.

Next, we turn to the neutrino production $\tilde{\chi}^0_2 \rightarrow
\tilde{\chi}^0_1 \nu \bar{\nu}$. The Higgs exchange does not contribute. For
the partial width via $Z$ exchange, we get the same result as for charged
leptons with an extra factor 2 from the $Z \nu \bar{\nu}$ vertex. As the
sneutrino is a "left" type sfermion, the vertex factor required for the
sneutrino exchange is slightly different: \bea \frac{1}{\sqrt{2}}(g_2N_{11} -
g_1N_{12}) \simeq .17 \lambda \left(\frac{M_Z}{M_1}\right)^2 . \eea The partial
width reads \bea \Gamma(\tilde{\chi}^0_2 \stackrel{\tilde{\nu}}\longrightarrow
\tilde{\chi}^0_1 \nu \bar{\nu}) & \simeq & 10^{-6} \lambda^2 M_1
\left(\frac{M_Z}{m_{\tilde{\nu}}}\right)^4 I_{\tilde{\nu}}
(\eta,\omega_{\tilde{\nu}}) \nn \\ & \gsim & 10^{-8} \lambda^2 M_1 . \eea
Although sneutrinos can be rather heavy ($\sim$~150~GeV for $M_1=M_Z$), this
contribution always remains larger than the one from $Z$ exchange. Thus, the
virtual sneutrino exchange gives the main contribution to the $\tilde{\chi}^0_2
\rightarrow \tilde{\chi}^0_1 \nu \bar{\nu}$ channel and the computation of
interference terms is not needed.

Finally we consider the decay into two jets $\tilde{\chi}^0_2 \rightarrow
\tilde{\chi}^0_1 q \bar{q}$. The top is too heavy to be produced. The partial
width via virtual $Z$ exchange is of the same order as in the case of leptons
(with an extra color factor $N_q=3$ and slightly different $Z f \bar{f}$
couplings), whereas the squark exchange is strongly suppressed because squarks
are always rather heavy. As for charged leptons, the virtual Higgs exchange
plays no role. Hence, the virtual $Z$ exchange is the only important
contribution to the $\tilde{\chi}^0_2 \rightarrow \tilde{\chi}^0_1 q \bar{q}$
partial width, which therefore is always small compared to the partial width
into two leptons via slepton exchange.

With the approximate expressions for the three body decays $\tilde{\chi}^0_2
\rightarrow \tilde{\chi}^0_1 l^+ l^-$, $\tilde{\chi}^0_2 \rightarrow
\tilde{\chi}^0_1 \nu \bar{\nu}$ and $\tilde{\chi}^0_2 \rightarrow
\tilde{\chi}^0_1 q \bar{q}$ at hand, we turn now to the "exact" numerical
results. In figs.~\ref{figbrl}-\ref{figbrq} we present our numerical results
for the branching ratios of the three body decays $\tilde{\chi}^0_2 \rightarrow
\tilde{\chi}^0_1 l^+ l^-$, $\tilde{\chi}^0_2 \rightarrow \tilde{\chi}^0_1 \nu
\bar{\nu}$ and $\tilde{\chi}^0_2 \rightarrow \tilde{\chi}^0_1 q \bar{q}$
($q=u,d,c,s,b$) respectively, for $\sim 10^4$ points in the parameter space
described in sec.~2. Here we used exact expressions for the mixing factors, the
phase space integrals and we included  all contributions to a given final
state. 

 From the previous discussion, the branching ratios do not depend on $\lambda$,
since all the partial widths are proportional to $\lambda^2$, but essentially
on the bino mass: For small values of $m_{\chi_2^0}$ ($\sim$~30~GeV),
sneutrinos are lighter than charged sleptons ($\sim$~45~GeV and $\sim$~60~GeV
respectively), therefore the main contribution to the total decay width is the
neutrino production via virtual sneutrino exchange (fig.~\ref{figbrn}). As the
bino mass increases, the sneutrino mass gets larger than the charged slepton
mass. The dominant process is then the charged lepton production via virtual
slepton exchange (fig.~\ref{figbrl}) (except for a small domain in the
parameter space on which we shall come back in the next paragraph). As
advertised earlier, the jet production (fig.~\ref{figbrq}) is always small. In
the domain of large bino masses, where sleptons are also relatively heavy, the
quark production via virtual $Z$ exchange can contribute up to $\sim$ 20\% to
the total width. Genuinely we have \bea 10^{-4} \lsim
\frac{\Gamma(\tilde{\chi}^0_2 \rightarrow \tilde{\chi}^0_1 q
\bar{q})}{\Gamma(\tilde{\chi}^0_2 \rightarrow \tilde{\chi}^0_1 l^+ l^-)} \lsim
10^{-1} . \eea

Let us now study the two body decay into a real Higgs boson of
fig.~\ref{figprod}, starting again with approximate analytic expressions. The
lightest non-singlet scalar (which is mainly $H_{1R}$) is too heavy to be
produced on shell. As already remarked in (\ref{above2}), one gets for the
bino-singlino-singlet Higgs scalar vertex factor \bea Q_{112} \sim \lambda^2 ,
\eea so that the partial width (\ref{DWprod}) \cite{neu1} approximately reads
\bea \Gamma(\tilde{\chi}^0_2 \rightarrow \tilde{\chi}^0_1 S_R) \sim \lambda^4
M_1 . \label{above3} \eea For small values of $\lambda$, this is completely
negligible compared to the three body decay rates. Yet, (\ref{above3}) involves
no tiny numerical factor stemming from virtually exchanged particles. Hence, if
$\lambda$ is not too small, real Higgs singlet production can even dominate the
total decay width. However, as emphasized in the previous section, the masses
of $\tilde S$ and $S_R$ are roughly proportional to $\lambda^{.5 \pm .3}M_0$.
Therefore, if $\lambda$ is too large, the singlet Higgs scalar and the singlino
become too heavy to be produced on shell in the bino decay and this channel is
kinematically forbidden. 

The numerical results are displayed in fig.~\ref{figbrs}. They are in good
agreement with our approximations and one finds that for a small window in
$\lambda$, $\lambda \simeq 10^{-3}$, the branching ratio of this process can
reach 90\%. In the same window, we could have the two body decay with a real
singlet Higgs pseudoscalar $\tilde{\chi}^0_2 \rightarrow \tilde{\chi}^0_1 S_I$.
However, since the pseudoscalar singlet is always heavier than the scalar, this
contribution remains small ($\lsim$ 5\%). If $\lambda \gsim 2.10^{-3}$, the
emitted Higgs singlet is heavy enough to decay into $b \bar{b}$, which is then
the main final state. For smaller values of $\lambda$, this channel is
kinematically closed. Depending on the singlet mass, the $\tau^+ \tau^-$/$c
\bar{c}$ channels are then favored. Smaller singlet masses correspond to
smaller values of $\lambda$, in which case the real Higgs singlet production is
negligible.

Finally, we turn to the radiative decay $\tilde{\chi}^0_2 \rightarrow
\tilde{\chi}^0_1 \gamma$. A complete calculation involves loops with fermions +
sfermions and charginos + $W$, charged Higgs and Goldstone bosons (depending on
the gauge choice) \cite{HaWy}. The corresponding contributions decrease with
increasing masses of the particles inside the loops. In the following analytic
approximation, we then only consider the dominant diagram, involving the
lightest particles in the loops, namely the "right" type charged sleptons
(fig.~\ref{figrad}). However, it is worth being stressed that we performed a
complete numerical analysis, including all the loops mentioned above with the
correct chargino and stop mass eigenstates \cite{unpub}. The effective coupling
(\ref{coup}) for three degenerate $\tilde{l}_R$ loops is given by \bea
g_{\gamma} & = & \frac{3eg_1^2N_{12}}{16\pi^2}
I_{\gamma}(\eta,\omega_{\tilde{l}_R}) \nn \\ & \simeq & 2.10^{-4} \lambda
\left(\frac{M_Z}{M_1}\right)^2 I_{\gamma}(\eta,\omega_{\tilde{}_R}) \eea where
$I_{\gamma}$, defined in eq.~(\ref{Irad}), is of $O(10^{-1})$ if $|\eta|
\not\rightarrow 1$. The partial width (\ref{DWrad}) then reads \bea
\Gamma(\tilde{\chi}^0_2 \rightarrow \tilde{\chi}^0_1 \gamma) \sim 10^{-11}
\lambda^2 \frac{M_Z^4}{M_1^3} . \eea Even for small values of $M_1$, this is
totally negligible compared to the three body decay rates. This is not
surprising since it is a contribution of higher order in perturbation theory.
Note that there is no "dynamical enhancement" mechanism for this channel in our
model as it can appear in the MSSM under special assumptions \cite{HaWy,rad}.
However, there could be some "kinematical enhancement":

Up to now, we have assumed $|\eta| \ll 1$ (i.e. very light singlino) in all our
analytic approximations. What happens for $|\eta| \rightarrow 1$ ? On the one
hand, all the three body decay phase space integrals (\ref{IZ}), (\ref{Ia}) and
(\ref{Is}) tend towards 0. As it has already been mentioned elsewhere
\cite{HaWy,rad}, one can check that they are all of order \bea I(\eta,\omega)
\stackrel{|\eta| \rightarrow 1} \sim (1-|\eta|)^5 . \eea Hence, all the
$\tilde{\chi}^0_2 \rightarrow \tilde{\chi}^0_1 f \bar{f}$ channels are {\it
equally} suppressed. Furthermore, since in this case the singlino mass is close
to the bino mass, the two body decay with a real singlet Higgs boson is
kinematically forbidden.

On the other hand, it is well known that the radiative decay $\tilde{\chi}^0_2
\rightarrow \tilde{\chi}^0_1 \gamma$ is usually less suppressed for $|\eta|
\rightarrow 1$. One can expand the loop integral (\ref{Irad}) around $\eta =
\pm 1$: \bea I_{\gamma}(\eta,\omega_{\tilde{l}_R}) & \stackrel{\eta \rightarrow
1}\sim & (1-\eta)^5 \nn \\ & \stackrel{\eta \rightarrow -1}\sim & (1+\eta)^3 .
\eea Therefore, the radiative decay gives the main contribution to the total
decay width for $\eta \rightarrow -1$, but not for $\eta \rightarrow 1$. This
phenomenon has not been observed before in the context of radiative neutralino
decay. A similar effect exists for the neutrinoless double beta decay where the
result depends on the relative sign of the Majorana neutrino masses
\cite{2beta}. This rough estimate correctly fits our numerical results for the
branching ratio $Br(\tilde{\chi}^0_2 \rightarrow \tilde{\chi}^0_1 \gamma)$,
shown in fig.~\ref{figbrp}. Actually, one finds that the main contributions to
the radiative decay are the charged lepton/"right" slepton loops, the
top/lighter stop loops and the lighter chargino/$W$ loops \cite{unpub}.
Interferences between chargino and sfermion loops being destructive, this leads
to even smaller branching ratios for the radiative decay, and this channel is
kinematically enhanced only for a few points in the parameter space
corresponding to $\eta \sim -1$.

To conclude, we give in fig.~\ref{figtot} the total width of the bino to
singlino transition, including all the contributions discussed above, as a
function of $\lambda$. The $\lambda^2$ dependence is manifest for very small
values of $\lambda$ where the singlino is always very light. As $\lambda$
increases, the singlino mass can take non negligible values, in which case the
bino decay is kinematically suppressed. These cases correspond to the points in
fig.~\ref{figtot} below the "fat" diagonal line. From the total width it is
straightforward to compute the bino lifetime: \bea \tau_{\tilde{\chi}^0_2} =
\frac{\hbar} {\Gamma(\tilde{\chi}^0_2 \rightarrow \tilde{\chi}^0_1 X)} =
\frac{6.58\,10^{-25} \mbox{GeV.s}} {\Gamma(\tilde{\chi}^0_2 \rightarrow
\tilde{\chi}^0_1 X)} . \eea For an energetic bino, the length of flight in the
lab system is given by \bea l_{\tilde{\chi}^0_2} = \sqrt{\gamma^2-1} \, c
\tau_{\tilde{\chi}^0_2} \simeq \frac{\hbar c} {\Gamma(\tilde{\chi}^0_2
\rightarrow \tilde{\chi}^0_1 X)} \simeq \frac{1.97\,10^{-16} \mbox{GeV.m}}
{\Gamma(\tilde{\chi}^0_2 \rightarrow \tilde{\chi}^0_1 X)} \eea One can then
simply read off $l_{\tilde{\chi}^0_2}$ from fig.~\ref{figtot}. For
$\Gamma(\tilde{\chi}^0_2 \rightarrow \tilde{\chi}^0_1 X) \lsim 10^{-16}$~GeV
(which corresponds to $\lambda \lsim 5.10^{-6}$ or strong kinematical
suppression), the bino escapes the detector, and the signature is the same as
in the MSSM. In the other cases, we obtain the expected additional cascade,
with a macroscopically displaced vertex ($l_{\tilde{\chi}^0_2} > 1$~mm) for
$\Gamma(\tilde{\chi}^0_2 \rightarrow \tilde{\chi}^0_1 X) \lsim 5.10^{-13}$~GeV.

\section{Conclusions and Outlook}

The purpose of the present paper was the calculation of the NLSP partial and
total decay widths in the (M+1)SSM, in the case where the LSP is a singlino and
sparticle production is kinematically allowed at LEP2. Then, the NLSP is
essentially a bino and the bino to singlino transition appears at the end of
all sparticle decay chains. (On the other hand, if the singlino is {\it not}
the LSP, it will be nearly impossible to produce neither the singlino nor the
Higgs singlet in collider experiments, since the singlet sector is always
almost decoupled from the rest of the theory. The (M+1)SSM is then very
difficult to disentangle from the MSSM.)

We worked in the context of the constrained (M+1)SSM, with universal boundary
conditions for the soft terms at the GUT scale. The essential features of this
scenario have been discussed, using analytic approximations, in section 3. We
have seen that, while the singlino LSP scenario is not a necessary consequence
of the model, it corresponds to a natural "gaugino dominated" scenario, gaugino
masses being the largest soft terms. Furthermore, the singlino is automatically
the LSP for very small Yukawa couplings $\lambda$ and $\kappa$.  

The gross features of the bino decay widths are easy to understand using
analytic approximations. In section 4 we have presented and discussed such
approximate analytic expressions for the partial decay widths, which is in
good  agreement  with the results obtained numerically (without the
corresponding approximations). The numeric results have been presented in the
form of the figs.~\ref{figrad}-\ref{figtot}. (The corresponding detailed
formulae are given in the appendices.)

Let us summarize the behavior of the total decay width $\Gamma(\tilde{\chi}^0_2
\rightarrow \tilde{\chi}^0_1 X)$. In principle, the bino partial and total
decay widths can vary over many orders of magnitude, depending on the Yukawa
coupling $\lambda$, the bino mass $m_{\chi_2^0} \sim M_1$ and the singlino to
bino mass ratio $\eta$ (with $\eta \rightarrow 0$ for $\lambda \rightarrow 0$).
Generally one has \bea \Gamma(\tilde{\chi}^0_2 \rightarrow \tilde{\chi}^0_1 X)
\lsim 10^{-6} \lambda^2 M_1. \label{above4} \eea For $|\eta| \ll 1$, the
inequality (\ref{above4}) can roughly be replaced by an equality. For $|\eta|
\rightarrow 1$, however, $\Gamma(\tilde{\chi}^0_2 \rightarrow \tilde{\chi}^0_1
X)$ can be considerably smaller than the right hand side of (\ref{above4})
because of kinematical suppression. The allowed range of the total decay width
as a function of $\lambda$ is displayed in fig.~\ref{figtot}, and the
corresponding bino lifetime can lead to macroscopically displaced vertices. For
very small values of $\Gamma(\tilde{\chi}^0_2 \rightarrow \tilde{\chi}^0_1 X)$,
the bino may even decay outside the detector (in which case it imitates the
true LSP of the MSSM). However, this requires tiny Yukawa couplings ($\lambda
\lsim 5.10^{-6}$) or strong kinematical suppression. Such scenarios could be
probed by the same kind of apparatus as the slow neutralino to gravitino
transition in the context of Gauge Mediated Supersymmetry Breaking models
\cite{MaOr}.

For a given value of $\Gamma(\tilde{\chi}^0_2 \rightarrow \tilde{\chi}^0_1 X)$,
the branching ratios of the bino still vary essentially with the bino and
singlino masses. In most of the parameter space, the three body decay
(fig.~\ref{figexch}) dominates, and the relevant final states are $\nu
\bar{\nu}$, $l^+l^-$ or $q \bar{q}$ ($q=u,d,c,s,b$) and missing energy. For
small values of the bino mass ($\sim$~30~GeV), the $\nu \bar{\nu}$ channel
dominates (fig~\ref{figbrn}). Hence, the bino decays invisibly and its
signature is just missing energy as for the true LSP of the MSSM. However, this
channel never exceeds 90\%, the remaining 10\% corresponding to the visible
$l^+l^-$ channel (fig.~\ref{figbrl}). For larger values of the bino mass (up to
$M_Z$), on the other hand, the invisible final state $\nu \bar{\nu}$ becomes
less important and the charged lepton channel contributes up to 100\%. The
partial width into $q \bar{q}$ is always small compared to the partial width
into $l^+l^-$, and we expect at most one jet event for ten charged lepton
events. The characteristic signature for sparticle production would then be
lepton events with high multiplicity (at least four, in $e^+e^- \rightarrow
\tilde{\chi}_2^0 \tilde{\chi}_2^0$) plus missing energy, eventually with
displaced vertices.

However, in the window $10^{-3} \lsim \lambda \lsim 10^{-2}$, the two body
decay $\tilde{\chi}^0_2 \rightarrow \tilde{\chi}^0_1 S_1$ dominates, if
kinematically allowed (fig.~\ref{figbrs}). $S_1$ is then essentially the Higgs
singlet with a mass varying between 3 and 35~GeV. If its mass is larger than
$\sim$~10~GeV, $S_1$ decays into $b \bar{b}$ (with a branching ratio of $\sim$
90\%); otherwise, $\tau^+ \tau^-$/$c \bar{c}$ are favored. (For such values of
$\lambda$, the bino and the real Higgs singlet will have short lifetimes.) The
relevant final state would be two $b$-jets (or eventually $\tau^+
\tau^-$/$c$-jets) with an invariant mass peaked below 35~GeV. Such processes
are totally exluded in the MSSM and would be a strong sign for the (M+1)SSM.

Finally, in the degenerate case $\eta \sim -1$, all the previous tree level
channels are kinematically suppressed, and the radiative decay
$\tilde{\chi}^0_2 \rightarrow \tilde{\chi}^0_1 \gamma$ dominates
(fig.~\ref{figbrp}). In such a scenario, the bino would be very long lived
($l_{\tilde{\chi}^0_2} \gsim 1$~m). (This corresponds, however, to a tiny
fraction of the parameter space.) In contrast to the MSSM \cite{rad}, a
dominance of the radiative decay is compatible with universal soft terms, and
the (rather disfavored) condition $\tan \beta \sim 1$ is not required in the
(M+1)SSM.

Herewith we have summarized our results, which have been obtained for
$m_{\chi_2^0} < M_Z$, the range accessible at LEP2. Let us, at the end, comment
on the case of a bino which is heavier than the $Z$. One should consider two
different regimes:

In the intermediate range $M_Z < m_{\chi_2^0} \lsim$~130~GeV, the main decay
mode becomes  $\tilde{\chi}^0_2 \rightarrow \tilde{\chi}^0_1 Z$, with
$\tilde{\chi}^0_1$ and $Z$ on shell. The total decay width is again
proportional to $\lambda^2$, though larger than in the case of the three body
decay (since no virtual particle needs to be exchanged). Hence, the bino would
not be too long lived in this case, except for extremely small values of
$\lambda$. The characteristic signature for this additionnal cascade is missing
energy plus the typical $Z$ decay products.

For a very heavy bino, $m_{\chi_2^0} \gsim$~130~GeV, one has $m_{\chi_2^0} >
m_{\tilde{l}_R}$ as already noticed in eq.~(\ref{slepbino}). The "right"
charged sleptons are hence light enough to be produced on shell, and the main
channel is $\tilde{\chi}^0_2 \rightarrow l \tilde{l}_R$. Since one needs at
least $m_{\chi_2^0} \gsim$~250~GeV in order to have $m_{\chi_2^0} -
m_{\tilde{l}_R} \gsim$~10~GeV, the emitted lepton is very soft, hence difficult
to detect. Yet, in this case, the true NLSP is the charged slepton and the
process of interest is $\tilde{l}_R \rightarrow l^\pm \tilde{\chi}^0_1$,
appearing at the end of all sparticle decay chains. Then one obtains a charged
slepton (eventually long lived, depending on $\lambda$) decaying into an
energetic lepton plus missing energy (the singlino). This case corresponds,
however, to a very heavy sparticle spectrum, disfavored by solutions to the
hierarchy problem.

\vspace{1cm} \noindent {\Large{\bf Acknowledgments}} \vspace{.5cm}

It is a pleasure to thank L. Duflot and C.A. Savoy for valuable discussions.

\newpage \noindent {\Large{\bf Appendices}} \appendix

\section{Neutralino Sector} \label{appneu}

The mass terms for the neutralinos are given by the following part of the
lagrangian: \bea {\cal L} & = & \frac{ig_1}{\sqrt{2}}(-h_1\lambda_1\psi_1 +
h_2\lambda_1\psi_2) + \frac{ig_2}{\sqrt{2}}(h_1\lambda_2^3\psi_1 -
h_2\lambda_2^3\psi_2) \nn \\ & & + \lambda s\psi_1\psi_2 + \lambda
h_1\psi_2\psi_s + \lambda h_2\psi_1\psi_s - \kappa s\psi_s\psi_s \nn \\ & & +
\frac{1}{2}M_1\lambda_1\lambda_1 + \frac{1}{2}M_2\lambda_2^3\lambda_2^3 +
\mbox{h.c.} \eea where the two component spinors $\lambda_1$, $\lambda_2^3$,
$\psi_1$, $\psi_2$ and $\psi_s$ are the supersymmetric partners of the $B$,
$W^3$, $H_1^0$, $H_2^0$ and $S$ bosons respectively. We introduce the 5
component neutralino vector \cite{HK}: \bea (\psi^0)^T = (-i\lambda_2^3,
-i\lambda_1,\psi_1 ,\psi_2, \psi_s) . \label{basis} \eea Then the mass terms
read \bea {\cal L} = -\frac{1}{2}(\psi^0)^TM^0(\psi^0) + \mbox{h.c.} \eea where
the (symmetric) neutralino mass matrix $M^0$ is given by \bea M^0 = \left(
\ba{ccccc} M_2 & 0 & \displaystyle{\frac{-g_2h_1}{\sqrt{2}}} &
\displaystyle{\frac{g_2h_2}{\sqrt{2}}} & 0 \\ & M_1 &
\displaystyle{\frac{g_1h_1}{\sqrt{2}}} &
\displaystyle{\frac{-g_1h_2}{\sqrt{2}}} & 0 \\ & & 0 & -\mu & -\lambda h_2 \\ &
& & 0 & -\lambda h_1 \\ & & & & 2\kappa s \ea \right) . \label{neumm} \eea

The physical mass eigenstates are obtained by diagonalizing ${\cal M}^0$ with a
unitary matrix $N$: \bea m_{\chi^0_i}\delta_{ij} = N_{im}N_{jn}M^0_{mn}
\label{defn} \eea $m_{\chi^0_i}$ being the mass eigenvalues in increasing order
of the neutralino states: \bea \chi_i^0 = N_{ij}\psi^0_j \quad i = 1, \ldots
,5. \label{neumix} \eea

We take $N$ real and orthogonal. Then some of the mass eigenvalues may be
negative. Finally, one obtains the proper 4 component neutralino eigenstates by
defining the Majorana spinors: \bea \tilde{\chi}_i^0 = \left( \ba{cc} \chi_i^0
\\ \bar{\chi}_i^0 \ea \right) \quad i=1, \ldots ,5. \eea

\section{Higgs Sector} \label{apphgs}

We give here the potential, the minimization equations and the mass matrix for
the neutral scalar Higgs without radiative corrections. The purpose of this
appendix is only to set up our conventions and to provide some guidelines in
order to make our analytic approximations easier to understand. A more complete
analysis of this sector can be found in \cite{Higgs}.

The scalar potential for the neutral Higgs fields is given by \bea V & = &
\lambda^2(|H_1^0|^2|S|^2 + |H_2^0|^2|S|^2 + |H_1^0|^2|H_2^0|^2) +
\kappa^2|S^2|^2 \nn \\ & & - \lambda\kappa (H_1^0H_2^0S^{*2} + \mbox{h.c.}) +
\frac{g^2}{4}(|H_1^0|^2 - |H_2^0|^2)^2 \nn \\ & & + m_1^2|H_1^0|^2 +
m_2^2|H_2^0|^2 + m_S^2|S|^2 \nn \\ & & - \lambda A_\lambda(H_1^0H_2^0S +
\mbox{h.c.}) +\frac{\kappa A_\kappa}{3}(S^3 + \mbox{h.c.}) \label{scalpot} \eea
where $g^2 = \frac{1}{2}(g_1^2+g_2^2)$. We split the Higgs fields into real and
imaginary parts:  \bea H_1^0 = h_1 + \frac{H_{1\,R}^0 + iH_{1\,I}^0}{\sqrt{2}}
, \quad H_2^0 = h_2 + \frac{H_{2\,R}^0 + iH_{2\,I}^0}{\sqrt{2}} , \quad S = s +
\frac{S_R + iS_I}{\sqrt{2}} . \label{hcomp} \eea

The conditions for extrema of (\ref{scalpot}) are \bea h_1(\lambda^2(h_2^2+s^2)
+ \frac{g^2}{2}(h_1^2-h_2^2) + m_1^2) - \lambda h_2s(A_\lambda + \kappa s) = 0
, \label{min1} \\ h_2(\lambda^2(h_1^2+s^2) - \frac{g^2}{2}(h_1^2-h_2^2) +
m_2^2) - \lambda h_1s(A_\lambda + \kappa s) = 0 , \label{min2}\\
s(\lambda^2(h_1^2+h_2^2) + 2\kappa^2s^2-2\lambda\kappa h_1h_2 + m_S^2) -
\lambda A_\lambda h_1h_2 + \kappa A_\kappa s^2 = 0 . \label{min3} \eea

After the elimination of $m_1^2$, $m_2^2$ and $m_S^2$ using
eq.~(\ref{min1}-\ref{min3}), the elements of the 3x3 symmetric mass matrix for
the CP-even scalars in the basis $(H_{1\,R}^0 , H_{2\,R}^0 , S_R)$ are \bea
M^2_{S,11} & = & g^2h_1^2 + \lambda s\frac{h_2}{h_1}(A_\lambda + \kappa s) , \\
M^2_{S,22} & = & g^2h_2^2 + \lambda s\frac{h_1}{h_2}(A_\lambda + \kappa s) , \\
M^2_{S,33} & = & \lambda A_\lambda\frac{h_1h_2}{s} + \kappa s(A_\kappa +
4\kappa s) , \label{ms33}\\ M^2_{S,12} & = & (2\lambda^2 - g^2)h_1h_2 - \lambda
s(A_\lambda + \kappa s) , \\ M^2_{S,13} & = & 2\lambda^2h_1s + \lambda
h_2(A_\lambda + 2\kappa s) , \\ M^2_{S,23} & = & 2\lambda^2h_2s + \lambda
h_1(A_\lambda + 2\kappa s) . \eea

Likewise, the elements of the 2x2 mass matrix for the CP-odd pseudoscalars in
the basis $(A^0,S_I)$ where the would-be Goldstone boson has been projected
out, read \bea M^2_{P,11} & = & \lambda s(A_\lambda + \kappa
s)\frac{h_1^2+h_2^2}{h_1h_2}, \\ M^2_{P,22} & = & \lambda(A_\lambda + 4\kappa
s)\frac{h_1h_2}{s} - 3\kappa A_\kappa s , \label{mp22}\\ M^2_{P,12} & = &
\lambda(A_\lambda - 2\kappa s)\sqrt{h_1^2+h_2^2}. \eea

The mass eigenstates of the scalars are denoted by $S_{a=1,2,3} (m_{S_1} \leq
m_{S_2} \leq m_{S_3})$ and those of the pseudoscalars by $P_{\alpha=1,2}
(m_{P_1} \leq m_{P_2})$.

\section{Approximate Results of the Integrated RGEs} \label{appRGE}

In this appendix we display some simple analytic results of the integrated RGEs
in the approximation where the dependence on all Yukawa couplings but the one
of the top quark, $h_t$, are neglected. Such solutions have been first
discussed in the MSSM framework in \cite{RGE1}. The Renormalization Group
Equations for the (M+1)SSM can be found in \cite{RGE2}. We assume universality
for the soft terms at the GUT scale and no flavor mixing. For a more complete
and general set of solutions, cf \cite{RGE3}.

First, let us define the parameter $\rho$ by \bea \rho =
\frac{h_t^2}{h_{crit}^2} \eea where $h_{crit}$ is the infra-red fixed point
solution for $h_t$, $h_{crit} \simeq 1.13$. We find from our numerical results
with a singlino LSP: $.7 \lsim \rho \lsim .8$.

The Yukawa couplings $\lambda$ and $\kappa$ are only slightly renormalized:
\bea \lambda \simeq \lambda_0 \quad , \quad \kappa \simeq \kappa_0 . \eea

The results for the soft trilinear terms read \bea A_\lambda & = &
A_0\left(1-\frac{\rho}{2}\right) + (1.11\rho-.59)M_0 , \label{RGEAl} \\
A_\kappa & = & A_0 . \label{RGEAk} \eea

Soft scalar masses are as follows \bea m_S^2 & = & m_0^2 , \label{RGEms} \\
m_1^2 & = & \left(1-\frac{3}{2}\rho\right)m_0^2 -
\frac{\rho(1-\rho)}{2}(A_0-2.22M_0)^2 \nn \\ & & + (.52-3.71\rho)M_0^2 ,
\label{RGEm1} \\ m_2^2 & = & m_0^2 + .52M_0^2 , \label{RGEm2} \\ m_L^2 & = &
m_0^2 + .52M_0^2 , \\ m_E^2 & = & m_0^2 + .15M_0^2 . \eea

Finally, we have the usual gaugino mass relations \bea M_2 = .82 M_0 , \qquad
M_1 = \frac{5}{3}\frac{g_1}{g_2}M_2 \simeq \frac{1}{2}M_2 = .41 M_0 . \eea

\section{Decay Widths} \label{appDW}

We first repeat some general features of the three body decay of
$\tilde{\chi}_2^0$ with mass $m_{\chi_2^0}$ and momentum $p_2$ into
$\tilde{\chi}_1^0$ with mass $m_{\chi_1^0}$ and momentum $p_1$ and two massless
fermions $f$ and $\bar{f}$ with momenta $k$ and $k'$ respectively.

Using the Mandelstam variables \bea s=(p_2-p_1)^2 , \qquad t=(p_2-k)^2 , \qquad
u=(p_1-k)^2 , \eea the differential decay width can be written as \bea d\Gamma
= \frac{N_f}{512 \pi^3 m_{\chi_2^0}^3}\sum_{spins}|{\cal M}|^2dudt \eea where
${\cal M}$ is the invariant amplitude for the processes under consideration and
$N_f$ is the color factor of the fermions. The different diagrams are shown in
fig.~\ref{figexch}. As seen in the main part of the paper, we only need the
integrated width for each process separately, without interference term (except
in the case of sfermion exchange, see below). The integration limits are \bea
\left\{ \ba{l} 0 \leq s \leq (|m_{\chi_2^0}| - |m_{\chi_1^0}|)^2 \\ t_{min,max}
= \frac{1}{2}\left(m_{\chi_2^0}^2 - m_{\chi_1^0}^2 - s \pm
\sqrt{\lambda(m_{\chi_2^0}^2,m_{\chi_1^0}^2,s)}\right) \\ u + t + s =
m_{\chi_2^0}^2 + m_{\chi_1^0}^2 \ea \right. \eea $\lambda$ being the usual
triangle function: \bea \lambda(a,b,c) = a^2+b^2+c^2-2ab-2ac-2bc . \eea

To begin with, let us consider the $Z$ exchange. We use the following
notations: \begin{itemize} \item The parameters in the $Z$-fermions coupling
are \bea L_f = T_{3f} - Q_f\sin^2\theta_W , \qquad R_f = - Q_f\sin^2\theta_W
\eea where $T_{3f}$ and $Q_f$ are the isospin and the charge of the fermion
respectively. \item The $Z$ propagator is given by  \bea D_Z(x) =
(x-M_Z^2+iM_Z\Gamma_Z)^{-1} . \eea \item The parameter in the $Z
\tilde{\chi}_1^0 \tilde{\chi}_2^0$ coupling is \bea O_{12} = N_{13}N_{23} -
N_{14}N_{24} \label{O12} \eea where $N_{ij}$ denote the mixing components of
the neutralinos as given by eq.~(\ref{neumix}). \end{itemize} The invariant
amplitude of the process is \cite{neu1} \bea \sum_{spins} |{\cal M}_Z|^2 & = &
4 g^4 (L_f^2+R_f^2) O_{12}^2|D_Z(s)|^2 \times \nn \\ & & \times \left(
(m_{\chi_2^0}^2 - t)(t - m_{\chi_1^0}^2) + t \leftrightarrow u +
2m_{\chi_2^0}m_{\chi_1^0}s \right) . \eea Hence, the partial width reads \bea
\Gamma(\tilde{\chi}_2^0 \stackrel{Z}\longrightarrow \tilde{\chi}_1^0 f \bar{f})
= \frac{N_fg^4(L_f^2+R_f^2)O_{12}^2}{16\pi^3}\frac{m_{\chi_2^0}^5}{M_Z^4}
I_Z(\eta,\omega_Z) . \label{DWZexch} \eea $I_Z$ is a phase space integral
depending on $\eta$ defined in eq.~(\ref{eta}) and $\omega_Z=m_{\chi_2^0}/M_Z$:
\bea I_Z(\eta,\omega_Z) = \frac{1}{3}
\int_{\frac{1-\eta^2}{2}}^{1-|\eta|}\!\!dz
\frac{\Delta(4z-1-2\eta+3\eta^2)(1+\eta-z)}{(\omega_Z^2(2z+\eta^2-1)-1)^2}
\label{IZ} \eea with $\Delta=\sqrt{(1-z)^2-\eta^2}$.

Next we turn to the Higgs exchange. We only display formulae for a CP-even
Higgs $S_a$. The $S_a \tilde{\chi}_1^0 \tilde{\chi}_2^0$ coupling reads \bea
Q_{a12} & = & \frac{\lambda}{\sqrt{2}}( S_{a1}(N_{15}N_{24}+N_{14}N_{25}) +
S_{a2}(N_{15}N_{23}+N_{13}N_{25}) \nn \\ & & +
S_{a3}(N_{13}N_{24}+N_{14}N_{23}) ) - \sqrt{2}\kappa S_{a3}N_{15}N_{25} \nn \\
& & + \frac{1}{2}(g_2N_{11}-g_1N_{12})(N_{23}S_{a1}-N_{24}S_{a2}) \nn \\ & & +
\frac{1}{2}(g_2N_{21}-g_1N_{22})(N_{13}S_{a1}-N_{14}S_{a2}) \label{Qa12} \eea
where $S_{ai}$ denote the mixing components of $S_a$ in the basis
(\ref{hcomp}). The Higgs-fermion coupling is \bea Q_{af} =
\frac{m_fS_{a1}}{\sqrt{2}\,h_1} & & \mbox{for an up-type quark,} \label{Qafu}
\\ Q_{af} = \frac{m_fS_{a2}}{\sqrt{2}\,h_2} & & \mbox{for a down-type quark or
a charged lepton} \label{Qafd} \eea where $m_f$ is the fermion mass. With the
Higgs propagator $D_a$, the invariant amplitude reads \bea \sum_{spins} |{\cal
M}_a|^2 = 4 Q_{a12}^2Q_{af}^2|D_a(s)|^2 (m_{\chi_2^0}^2 + m_{\chi_1^0}^2 +
m_{\chi_2^0}m_{\chi_1^0} - s)s , \eea leading to the following partial width:
\bea \Gamma(\tilde{\chi}_2^0 \stackrel{S_a}\longrightarrow \tilde{\chi}_1^0 f
\bar{f}) = \frac{N_f Q_{a12}^2Q_{af}^2}{16\pi^3}
\frac{m_{\chi_2^0}^5}{m_{S_a}^4} I_a(\eta,\omega_a) \label{DWHexch} \eea where
the phase space integral $I_a$ depends on $\eta$ and
$\omega_a=m_{S_a}/m_{\chi_2^0}$: \bea I_a(\eta,\omega_a) =
\int_{\frac{1-\eta^2}{2}}^{1-|\eta|}\!\!dz
\frac{\Delta\omega_a^4(2z-1+\eta^2)(1+\eta-z)}{(2z+\eta^2-\omega_a^2-1)^2}
\label{Ia} . \eea

Finally, we consider the sfermion exchange. Neglecting the fermion Yukawa
coupling, the $\tilde{\chi}_i^0f\tilde{f}_R$ vertex is \bea f_i =
g_1\sqrt{2}\,Q_fN_{i2} \label{f12R} , \eea and for the
$\tilde{\chi}_i^0f\tilde{f}_L$ vertex one gets \bea f_i =
\sqrt{2}\,(g_2T_{3f}N_{i1} - g_1(T_{3f}-Q_f)N_{i2}) \label{f12L} . \eea With
the sfermion propagator $D_{\tilde{f}}$, the invariant amplitude reads
\cite{neu1} \bea \sum_{spins} |{\cal M}_{\tilde{f}}|^2 & = &
f_1^2f_2^2\left((m_{\chi_2^0}^2 - t)(t - m_{\chi_1^0}^2)|D_{\tilde{f}}(t)|^2 +
t \leftrightarrow u \right. \nn \\ & & \left. +
2m_{\chi_2^0}m_{\chi_1^0}sD_{\tilde{f}}(t)D_{\tilde{f}}(u)\right) . \eea The
last term is an interference term. The partial width is then given by \bea
\Gamma(\tilde{\chi}_2^0 \stackrel{\tilde{f}}\longrightarrow \tilde{\chi}_1^0 f
\bar{f}) = \frac{N_f f_1^2f_2^2}{64\pi^3}
\frac{m_{\chi_2^0}^5}{m_{\tilde{f}}^4} I_{\tilde{f}}(\eta,\omega_{\tilde{f}}) .
\label{DWsexch} \eea $I_{\tilde{f}}(\eta,\omega_{\tilde{f}})$ is the phase
space integral, depending on $\eta$ and
$\omega_{\tilde{f}}=m_{\chi_2^0}/m_{\tilde{f}}$: \bea
I_{\tilde{f}}(\eta,\omega_{\tilde{f}}) & = & \frac{1}{4} \int_{\eta^2}^1\!\!dx
\frac{(1-x)^2(x-\eta^2)^2} {(\omega_{\tilde{f}}^2x-1)^2x} \nn \\ & + & \eta
\int_{\frac{1-\eta^2}{2}}^{1-|\eta|}\!\!dz\!
\int_{\frac{z-\Delta}{2}}^{\frac{z+\Delta}{2}}\!\!dx
\frac{2z+\eta^2-1}{(\omega_{\tilde{f}}^2(1-2x)-1)(\omega_{\tilde{f}}^2(1-2z+2x)-1)}
. \label{Is} \eea

Now, we turn to the two body decay. The decay rate for the on-shell scalar
Higgs production of fig.~\ref{figprod} reads \bea \Gamma(\tilde{\chi}_2^0
\rightarrow \tilde{\chi}_1^0 S_a) & = & \frac{Q_{a12}^2 m_{\chi_2^0}} {16\pi}
\sqrt{\lambda(1,\eta^2,\omega_a^2)} \, (1+2\eta+\eta^2-\omega_a^2) .
\label{DWprod} \eea

Finally, let us consider the radiative decay $\tilde{\chi}_2^0 \rightarrow
\tilde{\chi}_1^0 \gamma$. The analytic approximation used in the main part of
the paper includes only charged lepton/"right" slepton loops (for details, see
\cite{HaWy,unpub}). The different contributions are shown in
fig.~\ref{figrad}.  The decay width reads  \bea \Gamma(\tilde{\chi}_2^0
\rightarrow \tilde{\chi}_1^0 \gamma) =
\frac{g_{\chi\chi\gamma}^2(m_{\chi_2^0}^2 - m_{\chi_1^0}^2)^3} {8\pi
m_{\chi_2^0}^5} . \label{DWrad} \eea $g_{\chi\chi\gamma}$ is an effective
coupling: \bea g_{\chi\chi\gamma} = \frac{e}{16\pi^2}\sum_f
N_fQ_ff_1f_2I_{\gamma}(\eta,\omega_{\tilde{f}}) \label{coup} \eea where
$f_{1,2}$ are given in eq.~(\ref{f12R}) and \bea
I_{\gamma}(\eta,\omega_{\tilde{f}}) & = & \frac{1}{1-\eta^2} \int_0^1\!\!dx
\left( 1 + \eta +
\frac{1-\eta\omega_{\tilde{f}}^2x}{(1-\eta)\omega_{\tilde{f}}^2x} \log\left(
\frac{1-\omega_{\tilde{f}}^2x}{1-\eta^2\omega_{\tilde{f}}^2x} \right) \right) .
\label{Irad} \eea

\vspace{2cm} \section*{Figure Captions} \newcounter{fig} \begin{list}{\bf
figure \arabic{fig}:}{\usecounter{fig}}

\item Diagrams contributing to the three body decay $\tilde{\chi}_2^0
\rightarrow \tilde{\chi}_1^0 f \bar{f}$. \label{figexch}

\item Feynman graph for the two body decay $\tilde{\chi}_2^0 \rightarrow
\tilde{\chi}_1^0 S_a/P_\alpha$. \label{figprod}

\item Contributions to the radiative decay $\tilde{\chi}_2^0 \rightarrow
\tilde{\chi}_1^0 \gamma$ with only charged lepton/"right" slepton loops.
\label{figrad}

\item Branching ratio of the charged lepton production $\tilde{\chi}_2^0
\rightarrow \tilde{\chi}_1^0 l^+ l^-$ versus the Bino mass $m_{\chi_2^0}$.
\label{figbrl}

\item Branching ratio of the neutrino production $\tilde{\chi}_2^0 \rightarrow
\tilde{\chi}_1^0 \nu \bar{\nu}$ versus the Bino mass $m_{\chi_2^0}$.
\label{figbrn}

\item Branching ratio of the jet production $\tilde{\chi}_2^0 \rightarrow
\tilde{\chi}_1^0 q \bar{q}$ ($q = u,d,c,s,b$) versus the Bino mass
$m_{\chi_2^0}$. \label{figbrq}

\item Branching ratio of the real singlet Higgs scalar production
$\tilde{\chi}_2^0 \rightarrow \tilde{\chi}_1^0 S_1$ versus $\log(\lambda)$.
\label{figbrs}

\item Branching ratio of the radiative decay $\tilde{\chi}_2^0 \rightarrow
\tilde{\chi}_1^0 \gamma$ versus $\eta$. \label{figbrp}

\item Total decay width of the bino to singlino transition
$\log(\Gamma(\tilde{\chi}_2^0 \rightarrow \tilde{\chi}_1^0 X))$ versus
$\log(\lambda)$. \label{figtot}

\end{list}

\begin{figure}[p] \unitlength1cm \begin{picture}(12,20) \put(6.5,1.0){\bf
Figure 1} \put(-1,4){\epsffile{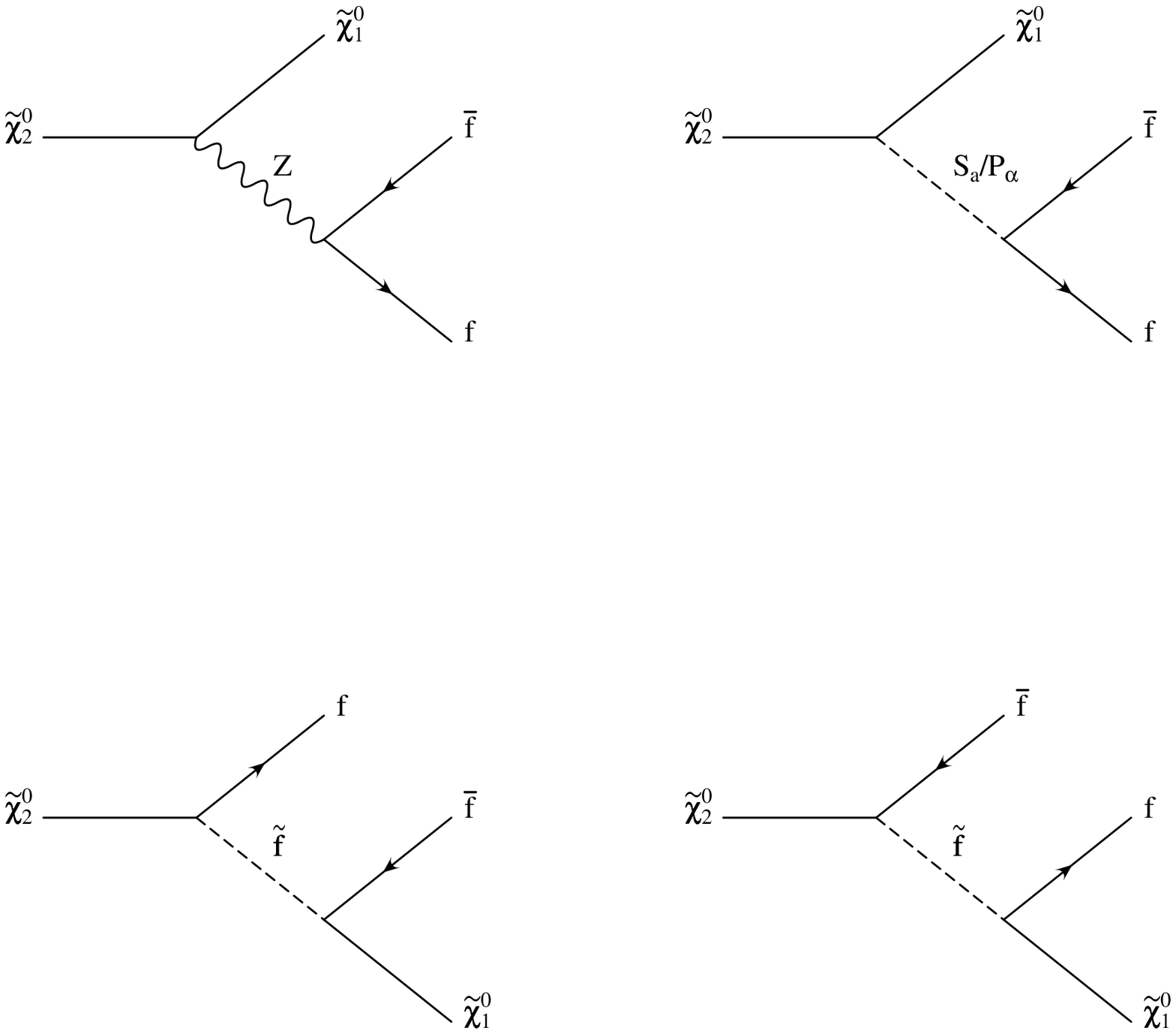}} \end{picture} \end{figure}

\begin{figure}[p] \unitlength1cm \begin{picture}(12,20) \put(6.5,12.0){\bf
Figure 2} \put(6.5,1.5){\bf Figure 3} \put(-1,4){\epsffile{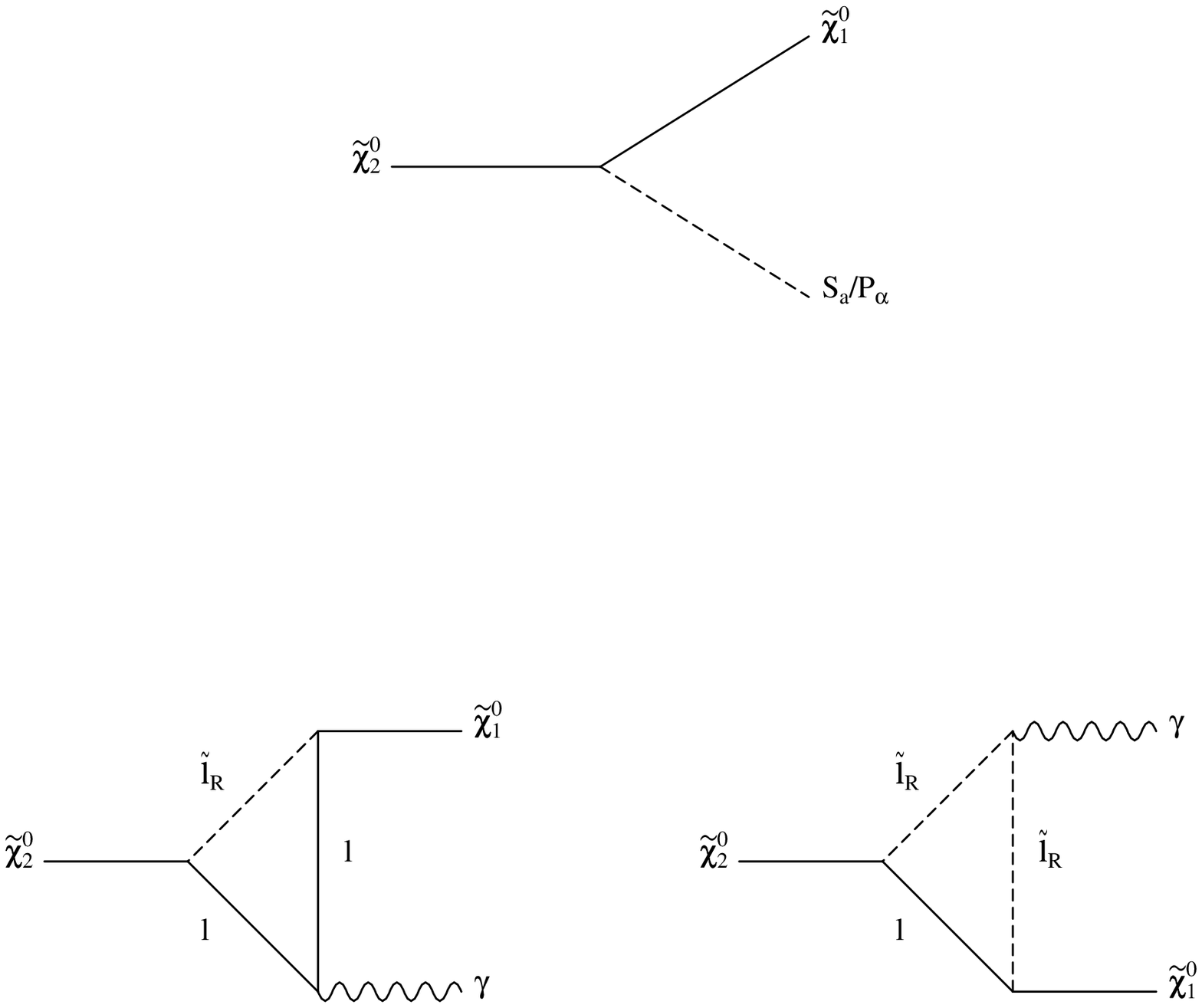}}
\end{picture} \end{figure}

\begin{figure}[p] \unitlength1cm \begin{picture}(12,20) \put(-1.8,20.5){${\bf
Br}(\tilde{\chi}_2^0 \rightarrow \tilde{\chi}_1^0 l^+ l^-)$}
\put(16.0,1.5){$\bf m_{\chi_2^0}$} \put(6.0,0.5){\bf Figure 4}
\put(-1.9,-4){\epsffile{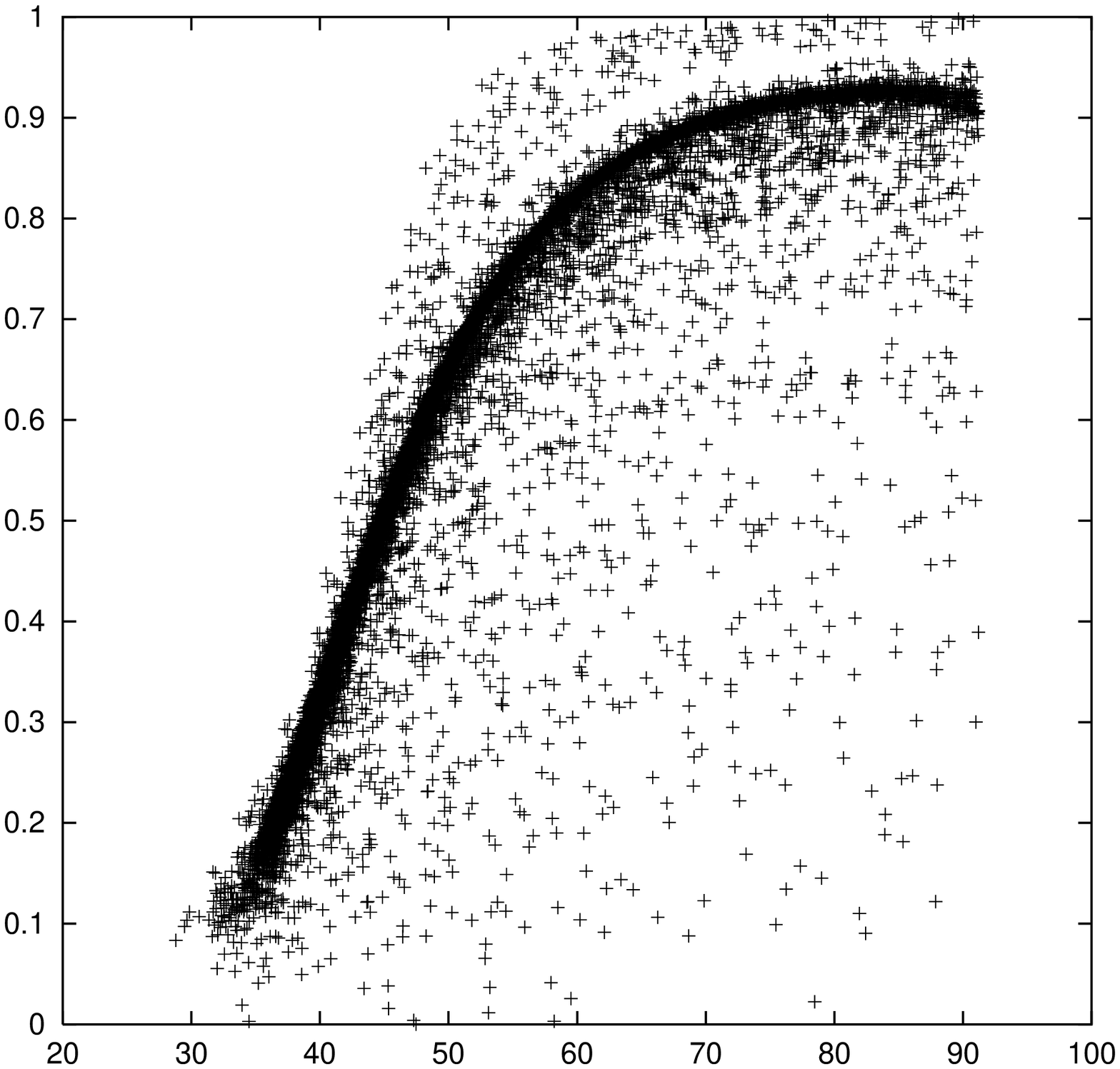}} \end{picture} \end{figure}

\begin{figure}[p] \unitlength1cm \begin{picture}(12,20) \put(-1.8,20.5){${\bf
Br}(\tilde{\chi}_2^0 \rightarrow \tilde{\chi}_1^0 \nu \bar{\nu})$}
\put(16.0,1.5){$\bf m_{\chi_2^0}$} \put(6.0,0.5){\bf Figure 5}
\put(-1.9,-4){\epsffile{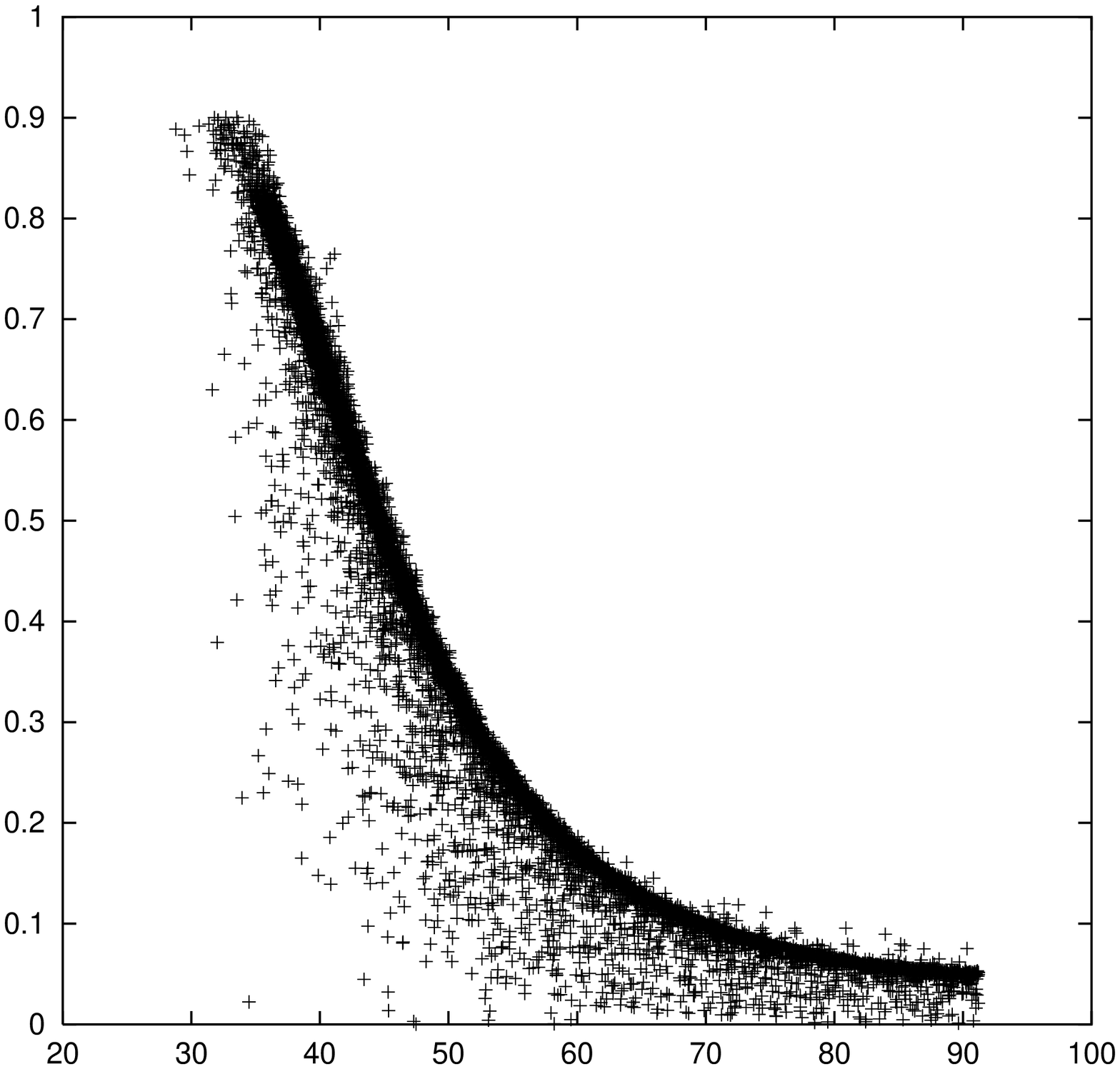}} \end{picture} \end{figure}

\begin{figure}[p] \unitlength1cm \begin{picture}(12,20) \put(-1.8,20.5){${\bf
{\displaystyle\sum_{q=u,d,c,s,b}} Br}(\tilde{\chi}_2^0 \rightarrow
\tilde{\chi}_1^0 q \bar{q})$} \put(16.0,1.5){$\bf m_{\chi_2^0}$}
\put(6.0,0.5){\bf Figure 6} \put(-1.9,-4){\epsffile{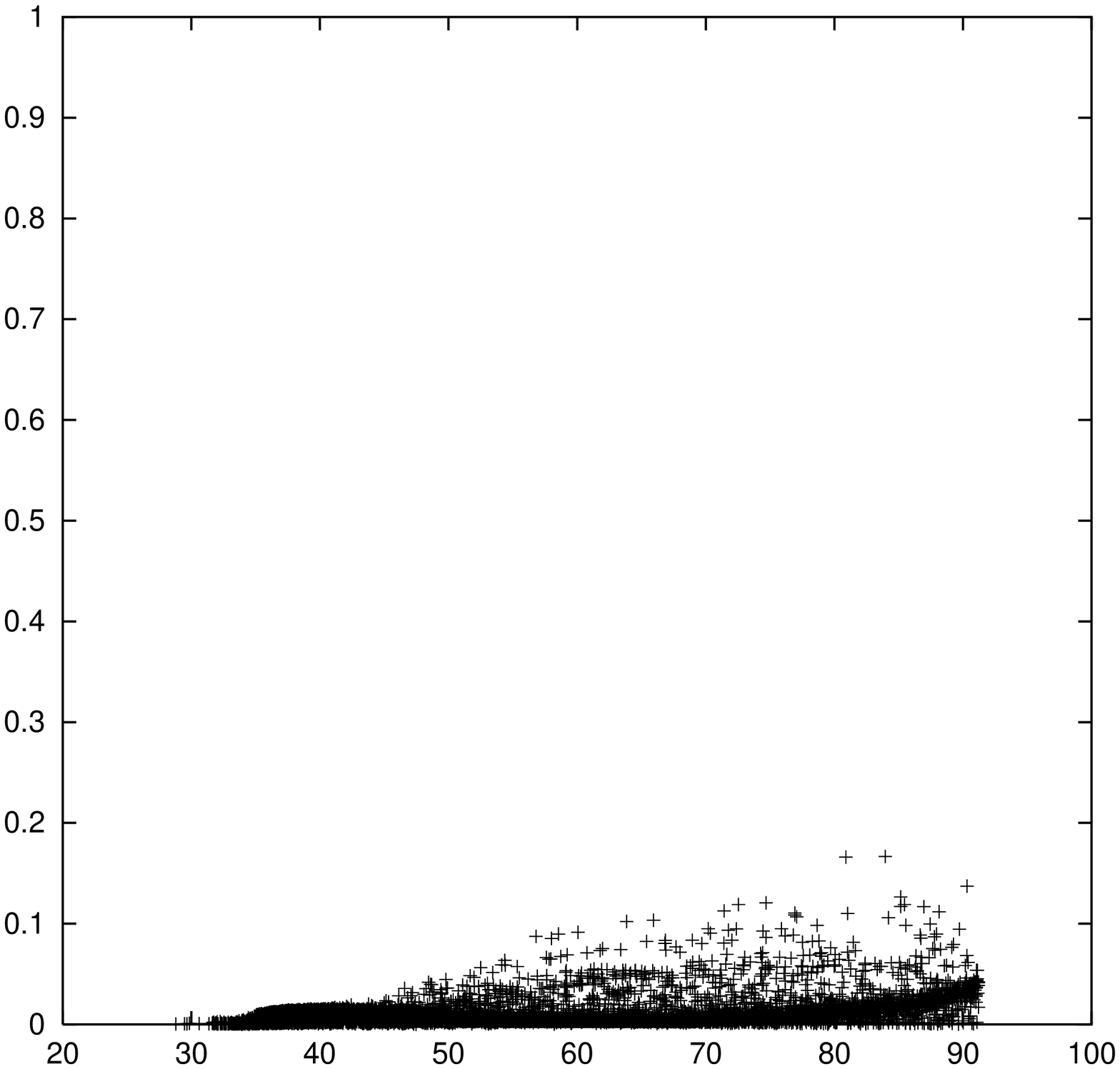}} \end{picture}
\end{figure}

\begin{figure}[p] \unitlength1cm \begin{picture}(12,20) \put(-1.8,20.5){${\bf
Br}(\tilde{\chi}_2^0 \rightarrow \tilde{\chi}_1^0 S_R)$} \put(16.0,1.5){$\bf
\log(\lambda)$} \put(6.0,0.5){\bf Figure 7} \put(-1.9,-4){\epsffile{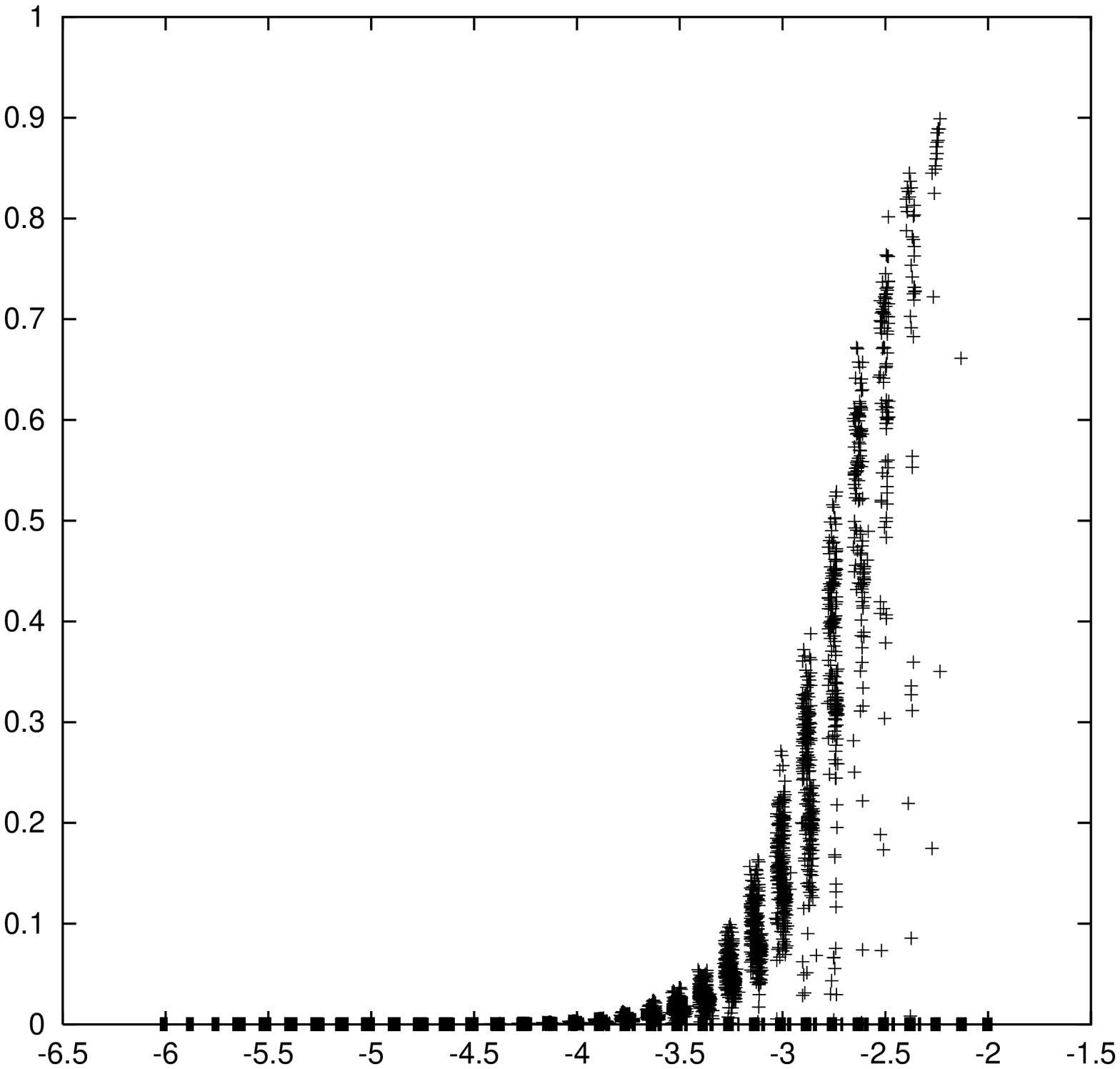}}
\end{picture} \end{figure}

\begin{figure}[p] \unitlength1cm \begin{picture}(12,20) \put(-1.8,20.5){${\bf
Br}(\tilde{\chi}_2^0 \rightarrow \tilde{\chi}_1^0 \gamma)$} \put(16.0,1.5){$\bf
\eta$} \put(6.0,0.5){\bf Figure 8} \put(-1.9,-4){\epsffile{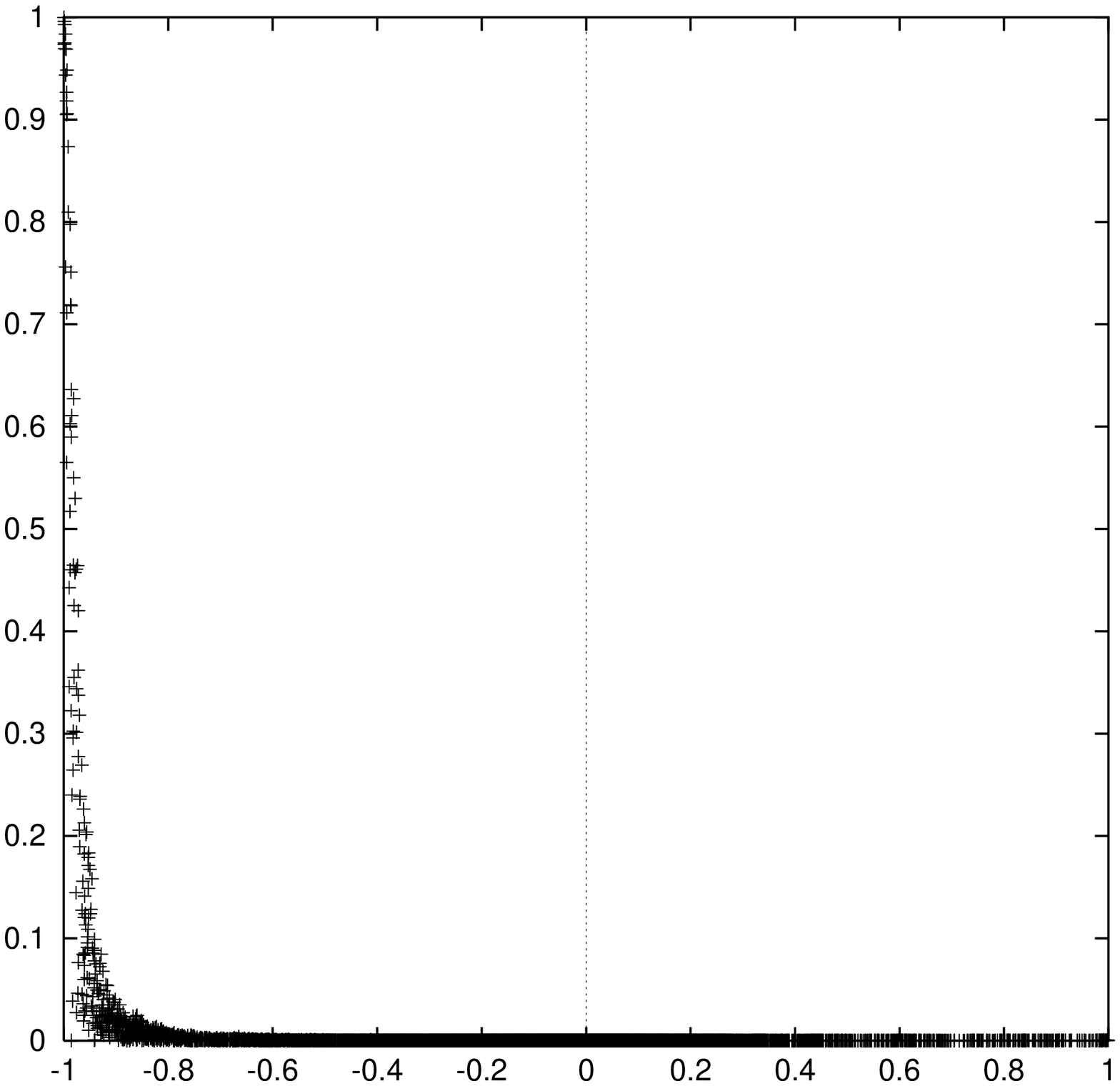}}
\end{picture} \end{figure}

\begin{figure}[p] \unitlength1cm \begin{picture}(12,20) \put(-1.8,20.5){$\bf
\log\left(\displaystyle\frac{\Gamma(\tilde{\chi}_2^0 \rightarrow
\tilde{\chi}_1^0 X)}{\rm 1~GeV}\right)$} \put(16.0,1.5){$\bf \log(\lambda)$}
\put(6.0,0.5){\bf Figure 9} \put(-1.9,-4){\epsffile{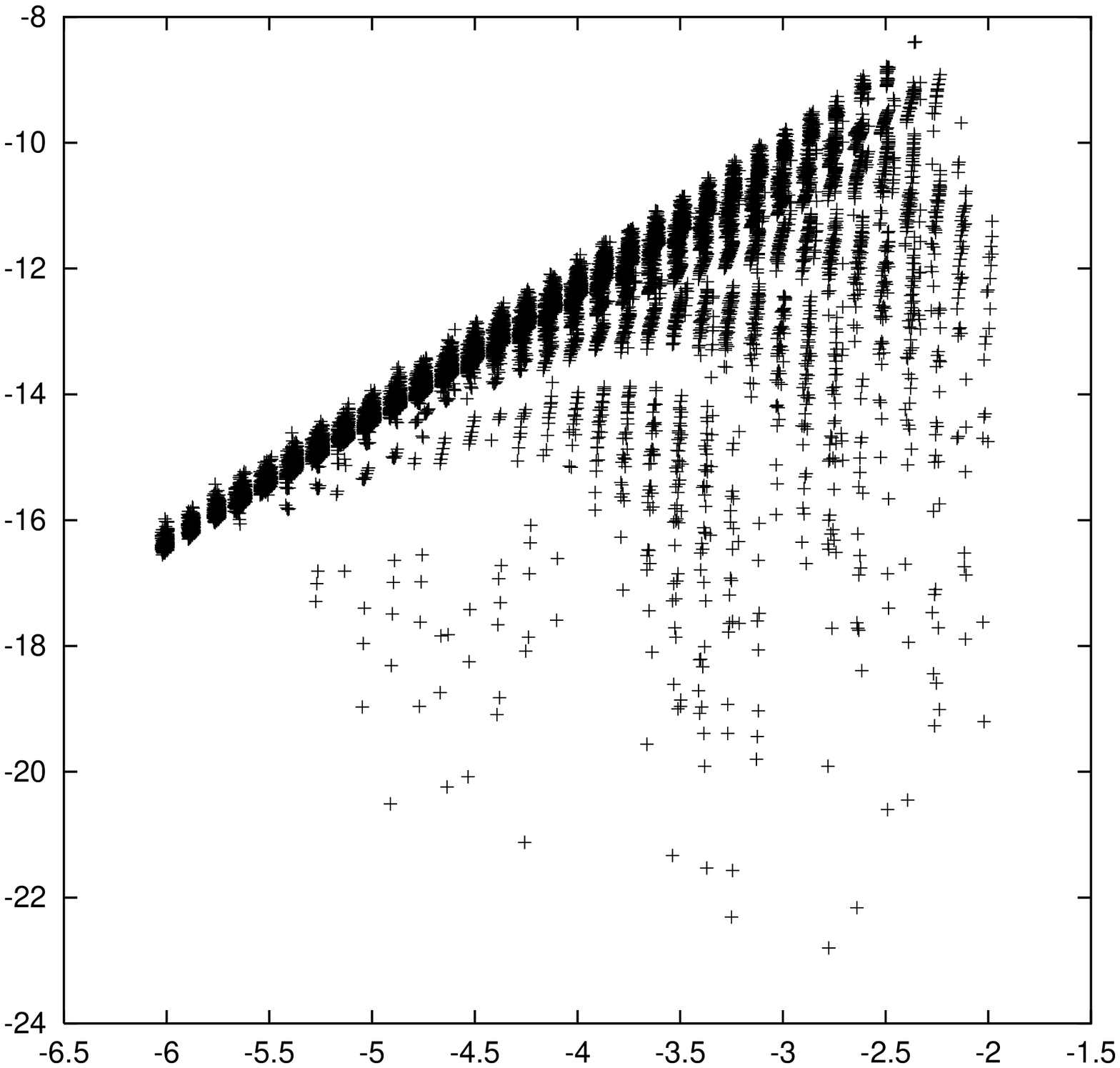}} \end{picture}
\end{figure}

\end{document}